\documentclass{emulateapj}
\usepackage{multirow}
\usepackage{natbib}     
\usepackage[flushleft]{threeparttable}

\usepackage{hyperref}

\begin{document}

\title{Modeling AGN Feedback in Cool-Core Clusters: The Balance between Heating and Cooling}
\author{Yuan Li and Greg L. Bryan}
\affil{Department of Astronomy, Columbia University, Pupin Physics Laboratories, New York, NY 10027}

\begin{abstract}
We study the long-term evolution of an idealized cool-core galaxy cluster under the influence of momentum-driven AGN feedback using three-dimensional high-resolution (60 pc) adaptive mesh refinement (AMR) simulations. The momentum-driven AGN feedback is modeled with a pair of (small-angle) precessing jets, and the jet power is calculated based on the accretion rate of the cold gas in the vicinity of the Supermassive Black Hole (SMBH). The ICM first cools into clumps along the propagation direction of the AGN jets. As the jet power increases, gas condensation occurs isotropically, forming spatially extended (up to a few tens kpc) structures that resemble the observed $\rm H\alpha$ filaments in Perseus and many other cool-core cluster. Jet heating elevates the gas entropy and cooling time, halting clump formation. The cold gas that is not accreted onto the SMBH settles into a rotating disk of $\sim 10^{11}$ M$_{\odot}$. The hot gas cools directly onto the cold disk while the SMBH accretes from the innermost region of the disk, powering the AGN that maintains a thermally balanced steady state for a few Gyr. The mass cooling rate averaged over 7 Gyr is $\sim 30$ M$_{\odot}$/yr, an order of magnitude lower than the classic cooling flow value (which we obtain in runs without the AGN). Medium resolution simulations produce similar results, but when the resolution is lower than 0.5 kpc, the cluster experiences cycles of gas condensation and AGN outbursts. Owing to its self-regulating mechanism, AGN feedback can successfully balance cooling with a wide range of model parameters. Besides suppressing cooling, our model produces cold structures in early stages (up to $\sim 2$ Gyr) that are in good agreement with the observations. However, the long-lived massive cold disk is unrealistic, suggesting that additional physical processes are still needed. 

\end{abstract}


\section{Introduction}
In most relaxed galaxy cluster, the temperature of the intracluster medium (ICM) decreases in the core \citep{Hudson2010}. The cooling time of the gas is much shorter than the Hubble time, allowing the system to reach a steady state where a cooling flow of 100s to 1000 M$_{\odot}$/yr was expected to develop (see \citet{Fabian1994} for a review). However, there is a lack of cooler gas at temperatures below 2-3 keV (or 1/2 to 1/3 of the virial temperature of the cluster) observed by Chandra and XMM-Newton \citep[e.g.,][]{Peterson2003, Peterson2006, Sanders2008}. This means that a classic cooling flow does not exist, and the cooling is balanced by some heating sources in these cool-core clusters. 

The proposed heating sources include conduction, star formation, AGN, cosmic rays and MHD waves \citep[e.g.][]{Narayan2003, Voit2005, Cattaneo2007, Falceta2010, Kunz2011}, among which AGN feedback is widely considered the most important and promising for the following reasons: (1) AGN heating is self-regulated, and can thus prevent over cooling or over heating; (2) AGN can be very energetic, and the observed AGN inflated bubbles have energy sufficient to balance the loss due to radiative cooling \citep[e.g.][]{McNamara2007}; (3) a strong correlation has been observed between the existence of an AGN and the cool-core status of the cluster \citep{Cavagnolo2008, Birzan2012}.

Besides the evidence of AGN activity, many cool-core clusters are also observed to harbor line-emitting gas \citep{Hu1985, Crawford1999, Donahue2000, Edge2001, Bregman2006, Mittal2012} and to form stars at a rate typically 1-2 orders of magnitude lower than the classic cooling flow rate \citep{Hicks2005, ODea2010}. These observations suggest the existence of a reduced cooling flow of some sort. In some cool-core clusters, the cold gas only exists in the nuclear region while in some others, the cold gas exhibit filamentary morphology that can extend to a few 10s kpc from the cluster center \citep{Conselice2001, McDonald10, McDonald11, McDonald12}. There is also a spatial correlation between the optical filaments and the enhanced soft X-ray and UV emission, indicating that the ICM cooling, AGN feedback, cold filaments and star formation are likely related \citep[e.g.][]{Fabian2003, Edwards2007}.

Previous simulations on AGN feedback in cool-core clusters have been focusing on how AGN deposits its energy and whether cooling can be balanced \citep[e.g.][]{Soker2001, Reynolds06, Gaspari2011}. It has been found that momentum driven AGN feedback can indeed suppress cooling over a few Gyr timescale \citep[e.g.][]{Cattaneo2007, Gaspari2011}, and \citet{Gaspari2012} has also seen the formation of multiphase gas, suggesting that momentum driven AGN feedback is a very successful model.

Prior to this work, we have carried out very high resolution simulations to study the onset of the cooling catastrophe in \citet{P1} (hereafter Paper I), and the formation of the cold filaments driven by the AGN feedback in \citet{PII} (hereafter Paper II). In this work, we focus on the long-term evolution of the cluster in the presence of momentum-driven AGN feedback. Besides the basic question ``can AGN feedback balance cooling'', we also try to address the following issues: (1) is cooling perfectly balanced by AGN feedback? (2) how does numerical resolution affect the results? (3) what are the required feedback parameters such as feedback efficiency? (4) does the model also produce cold gas distribution that is consistent with the observations? 

We describe the methodology of this work in Section~\ref{sec:method}. In Section~\ref{sec:results}, we present the main results of our standard simulation with $\sim 60$ pc resolution, and analyze how the energy is deposited, how efficient this heating mechanism is and how cooling is balanced by the heating. We also study how resolution and model parameters affect the results. In Section~\ref{sec:discussion}, we compare our results with the observations and previous simulations, and discuss the success and the caveats of our model. We summarize this work in Section~\ref{sec:conclusion}. As in Paper II, we refer to the diffuse ICM at temperatures above 1 keV as being ``hot'' and the condensed gas at temperatures around $\sim 10^4$ K as ``cold''.

\section{Methodology}
\label{sec:method}

We perform the simulations using Enzo \citep{Enzo}, a 3D AMR code. In the following, we summarize the refinement criteria of the simulations and the physics included in Section~\ref{sec:methodology_simulations}. In Section~\ref{sec:methodology_initial}, we describe the initial condition of the simulations and in Section~\ref{sec:methodology_jet}, the SMBH accretion and jet modeling.

\subsection{The Simulations}\label{sec:methodology_simulations}
The cooling function used in these simulations is the same as in Paper I \& II for radiative cooling, which is computed based on Table 4 from \citet{CoolingFunction} for gas with half-solar metallicity \citep{Metallicity} and a truncation temperature of $T_{\rm floor} \sim 10^4 K$. The refinement strategy is also the same, which is that a cell is refined whenever one of the following criteria is met: (1) the cell mass criterion -- if the mass of the gas in any cell exceeds $1/5$ of that in one cell of the root grid, (2) the cooling criterion -- when the ratio of the gas cooling time ($t_{\rm cool} = \frac{\frac{5}{2} n k_b T}{n^2 \Lambda(T)}$)to the sound-crossing time over the cell becomes too small, and (3) the Jeans length criterion -- when the cell size is larger than $1/4$ of the Jeans length. A detailed discussion of these criteria can be found in Paper I. We also add a nested static refine region around the jet launching region so that this region is always refined to the highest refinement level of the simulations.

All the simulations here have a box size of $L = 16$ Mpc and the number of root grids of $N_{\rm root} = 64$. Note that we use $N_{\rm root} = 256$ for the standard run in Paper II, but as we have shown in Paper I, changing $N_{\rm root}$ only mildly affects the temperature slope inside $r<10$ kpc during the pure cooling phase. Therefore, for the long term evolution when AGN feedback is included, we use $N_{\rm root} = 16$. In our standard run discussed in Section~\ref{sec:results_evolution}, the maximum refinement level is $l_{\rm max} = 12$, corresponding to the smallest cell size $\Delta x_{\rm min} = L / (N_{\rm root} 2^{l_{\rm max}}) \approx 60$ pc. The resolution of the other simulations is summarized in Table~\ref{table:parameters}, along with the important parameters of the model described in Section~\ref{sec:methodology_jet}. 

We do not include star formation, magnetic fields or heat conduction (yet). We will discuss the drawback in Section~\ref{sec:discussion_problems}. 

\begin{table*}
\caption{Simulations discussed in this work.
\label{table:parameters}}
\begin{center}
\begin{threeparttable}
\begin{tabular}{|l|l|l|l|l|}
\hline
Simulations    & $l_{max}$ & $f_{\rm kinetic}$ & $\epsilon$ & Figure and Section \\ \hline\hline 
Standard       & 12        & 0.5           & $0.1\%$    & Figure~\ref{fig:projections}, \ref{fig:cold_gas}, \ref{fig:kepler}, \ref{fig:mach}, \ref{fig:heating}, \ref{fig:Mdot}, \ref{fig:profile}, \ref{fig:sigma}, \ref{fig:M_T}, \ref{fig:X}, \ref{fig:M_r}   \\ \hline \hline 
\multirow{2}{*}{\begin{tabular}[c]{@{}c@{}}medium \\ resolution\end{tabular}} & 11        & 0.5           & $0.1\%$    & Section~\ref{sec:results_resolution}                 \\ \cline{2-4} & 10        & 0.5           & $0.1\%$    &    \\ \hline
\multirow{2}{*}{\begin{tabular}[c]{@{}c@{}}low \\ resolution\end{tabular}}    & 9         & 0.5           & $0.1\%$    & Section~\ref{sec:results_resolution}; Figure~\ref{fig:low_resolution}   
\\ \cline{2-4}  & 8         & 0.5           & $0.1\%$    &                    \\ \hline \hline 
high $f_{\rm kinetic}$    & 10        & 1             & $0.1\%$    & Section~\ref{sec:results_f}                   \\ \hline
low $f_{\rm kinetic}$   & 10        & 0.1           & $0.1\%$    & Section~\ref{sec:results_f}                   \\ \hline
high $\epsilon$   & 10        & 0.5           & $1\%$      & Section~\ref{sec:results_epsilon}; Figure~\ref{fig:high_epsilon} \\ \hline

\end{tabular}
\begin{tablenotes}[para,flushleft]
$l_{max}$ is the maximum refinement level, and $l_{max}=12$ gives a smallest cell size $\Delta x_{min} \sim 60$ pc. $f_{\rm{kinetic}}$ is the fraction of the jet energy in the form of kinetic energy, with the rest injected as thermal energy. $\epsilon$ is the feedback efficiency. Detailed description of these parameters can be found in Section~\ref{sec:methodology_jet}.
\end{tablenotes}
\end{threeparttable}
\end{center}
\end{table*}

\subsection{Cluster Initial Setup}\label{sec:methodology_initial}
The initial conditions are the same as in Paper II and are very similar to Paper I with a slight improvement. Our idealized galaxy cluster is built upon the observations of the Perseus cluster. We adopt the NFW parameters from \citet{Mathews}, where they assume that the gas is in hydrostatic equilibrium with the gravitational potential and fit the observed hydrostatic equilibrium gravity within $r < 300$ kpc with an NFW halo and a stellar component of the brightest cluster galaxy (BCG), NGC 1275:
\begin{equation}
M_*(r) = 
\frac{r^2}{G} \left[ \left( {r^{0.5975} \over 3.206 \times 10^{-7}}\right)^s
+ \left( {r^{1.849} \over 1.861\times 10^{-6}}\right)^s 
\right]^{-1/s}
\end{equation}
in cgs units with $s = 0.9$ and $r$ in kpc. The NFW halo parameters are: $M_{\rm vir} = 8.5 \times 10^{14}$ M$_{\odot}$, $r_{\rm vir} = 2.440$ Mpc and concentration $c=6.81$. Note that the NFW halo here includes both the dark matter halo and the ICM. We have shown in Paper I that the ICM contributes little to the total gravity, and therefore, for simplicity, we ignore the self-gravity of the ICM in our simulations presented here.

The observed electron density profile within $r < 300$ kpc follows:
\begin{equation}
n_e(r) = \frac{0.0192}{1 + \left(\frac{r}{18}\right)^3} + \frac{0.046}{\left[1 + \left(\frac{r}{57}\right)^2\right]^{1.8}}
+ \frac{0.0048}{\left[1 + \left(\frac{r}{200}\right)^2\right]^{0.87}} \rm{cm}^{-3}.
\end{equation}

And the observed azimuthally averaged temperature profile is:
\begin{equation}
T = 7 \; \frac{1 + (r/71)^3}{2.3 + (r/71)^3} \; \rm{keV}\;,
\end{equation}
where $r$ is the distance to the center of the cluster in kpc \citep{Churazov}.

At $r > 300$ kpc, due to the lack of observations, the temperature profile is computed with the universal formula found by  \citet{UniversalT} for cluster outskirts, normalized to match the observations at $r=300$ kpc:
\begin{equation}
T = 9.18 \times (1+(\frac{3\;r}{2\;r_{\rm vir}})^{-1.6})\; \rm{keV}\;.
\end{equation}

Given the NFW halo parameters and the temperature profile, we then compute the gas density and pressure profiles assuming hydrostatic equilibrium and the ideal gas law with an adiabatic index $\gamma = 5/3$, normalized such that the gas density is $15 \%$ of the total NFW density at the outskirt of the cluster. The main difference between the initial setup used here and in Paper I is how we model the gas properties at $r > 300$ kpc. In Paper I, we steepen the index of the observed gas density profile for larger radii, while here we adopt a universal temperature profile for the cluster outskirts. The resulting profiles are quite similar and since the cooling time of the gas at $r > 300$ kpc is longer than the Hubble time, the slight change of the initial condition has a negligible effect on the results. Another difference from Paper I is that in order to focus on the effect of AGN feedback here, we do not introduce any initial perturbation or rotation of the gas.

\subsection{Accretion and Jet Modeling}\label{sec:methodology_jet}  

The accretion and jet modeling is the same as in Paper II. Following \citet{Omma}, the jets in our simulations are launched along the z-axis from two circular planes perpendicular to the z-axis at a distance of $h_{jet}$ from the x-y plane.  Mass, momentum and thermal energy are added to the cells in the planes at each time step. During each time step $\Delta t$, the amount of mass added to the cells follows $\Delta m \propto e^{-r^2/2r^2_{\rm jet}}$ where r is the distance of the cell center from the z-axis. In the standard run, $h_{jet}$ is set to be 5 cell widths and $r_{jet}$ is 3. They are reduced to 2 and 1.5 cell widths respectively in the lower resolution runs to so that the physical size would increase less with lower resolution (and larger $\Delta x_{min}$). The effect of resolution is discussed in Section~\ref{sec:results_resolution}. $\Delta m$ is normalized such that the total outflow mass  $\int \Delta m$ during $\Delta t$ is equal to $\dot{M} \Delta t$, the total mass that is added to the accretion disk, which is not resolved in our simulations. The outflow rate is set to be equal to the accretion rate $\dot{M}$ because the difference between the two is the black hole growth rate, which should be negligible in radio mode AGN feedback considered here. We refer to the growth rate of the accretion disk as the accretion rate, and the accretion rate onto the SMBH as the ``SMBH growth rate'' through out the paper.

The accretion rate $\dot{M}$ is estimated by dividing $M_{cold}$, the total amount of cold gas in the vicinity of the SMBH, by $\tau$, a typical accretion time. The vicinity of the SMBH in our simulations is defined as a box of size $2\times r_{accretion}$ centered around the SMBH, with $r_{accretion} = 500$ pc in our standard simulation. The value of $r_{accretion}$ is chosen to be small enough to represent the sphere of influence of the SMBH, but large enough so that the accretion region is refined by at least a few cells. Thus for the two test runs with the lowest resolution, $r_{accretion}$ is increased to 1 kpc. The gas is considered ``cold'' when the temperature of the cell is lower than $T_{cold} = 3\times 10^4 K$. $\tau$ is set to be 5 Myr, which is roughly the average free-fall time of the gas within $r_{accretion}$. The simulation results are not sensitive to the exact choice of $T_{cold}$ and $\tau$.

After $\dot{M}$ is computed as each time step, the cold gas in the accretion region is accreted by removing a fraction of the mass of the cold cells. The total jet power is:
\begin{equation}\label{eq:Edot}
\dot{E} = \epsilon  \dot{M} c^2 \; ,
\end{equation}
where $\epsilon$ is the feedback efficiency with a typical value of $10^{-4}-10^{-2}$. Since the actually jet base is much closer to the SMBH than the jet launching planes in our simulations, some of the kinetic energy may have already thermalized. Therefore, the kinetic energy of the jets is only $\dot{E}_{\rm kinetic}=\dot{E}-\dot{E}_{\rm thermal}=f_{\rm kinetic}\dot{E}$. We will discuss the effect of varying feedback efficiency $\epsilon$ and kinetic fraction $f_{\rm kinetic}$ in Section~\ref{sec:results_other_simulations}. 

Given $\dot{M}$ and $\dot{E}_{\rm kinetic}$, the velocity of the injected material $v_{\rm jet}$ can be computed from $\dot{E}_{\rm kinetic}= \frac{1}{2}\dot{M}v^2_{\rm jet}$, and therefore, $v_{\rm jet}=\sqrt{2f\epsilon}c$. For $\epsilon=0.001$ and $f_{\rm kinetic}=0.5$, $v_{\rm jet} \approx 10^4$ km s$^{-1}$.

The evidence for the re-orientation of jets over $\sim 10s$ Myr timescales is seen in nearby cool-core clusters \citep{DunnPrecession, Babul2013}, but the physics of the jet re-orientation is beyond the scope of this study. To mimic the re-orientation of the jet, we make the jet precess at a small angle $\theta=0.15$ with a period of $\tau_p = 10$ Myr. This also help avoid the ``dentist drill'' effect with simple straight jet where the energy is channeled out of the cluster core seen in high resolution hydro simulations \citep[e.g.,][]{Reynolds06}. Varying $\tau_p$ or $\theta$ does not significantly change the results of our simulations. 

\section{Results}
\label{sec:results}
In this section, we present the main results of our simulations. We first describe the cluster evolution in our standard simulation in Section~\ref{sec:results_evolution}, and in Section~\ref{sec:results_heating_cooling}, we analyze how AGN deposits its energy and balances cooling over the course of a few Gyr. In Section~\ref{sec:results_other_simulations}, we look at the impact of resolution and model parameters on the simulation results.

\subsection{Cluster Evolution}\label{sec:results_evolution}
We present here the main results of our standard run over a few Gyr timescale. During the first 200 Myr, the temperature of the hot ICM in the core of the cluster slowly decreases and the density increases due to radiative cooling. A global cooling catastrophe first starts to happen in the very center ($r<100$ pc) of the cluster at $t\sim 200$ Myr, but no gas condensation is seen outside the central cooling region. The details of the cluster evolution during the pure cooling phase can be found in Paper I. 

\begin{figure*}
\begin{center}
\includegraphics[scale=.22,trim=0.8cm 2.2cm 0.8cm 0cm, clip=true]{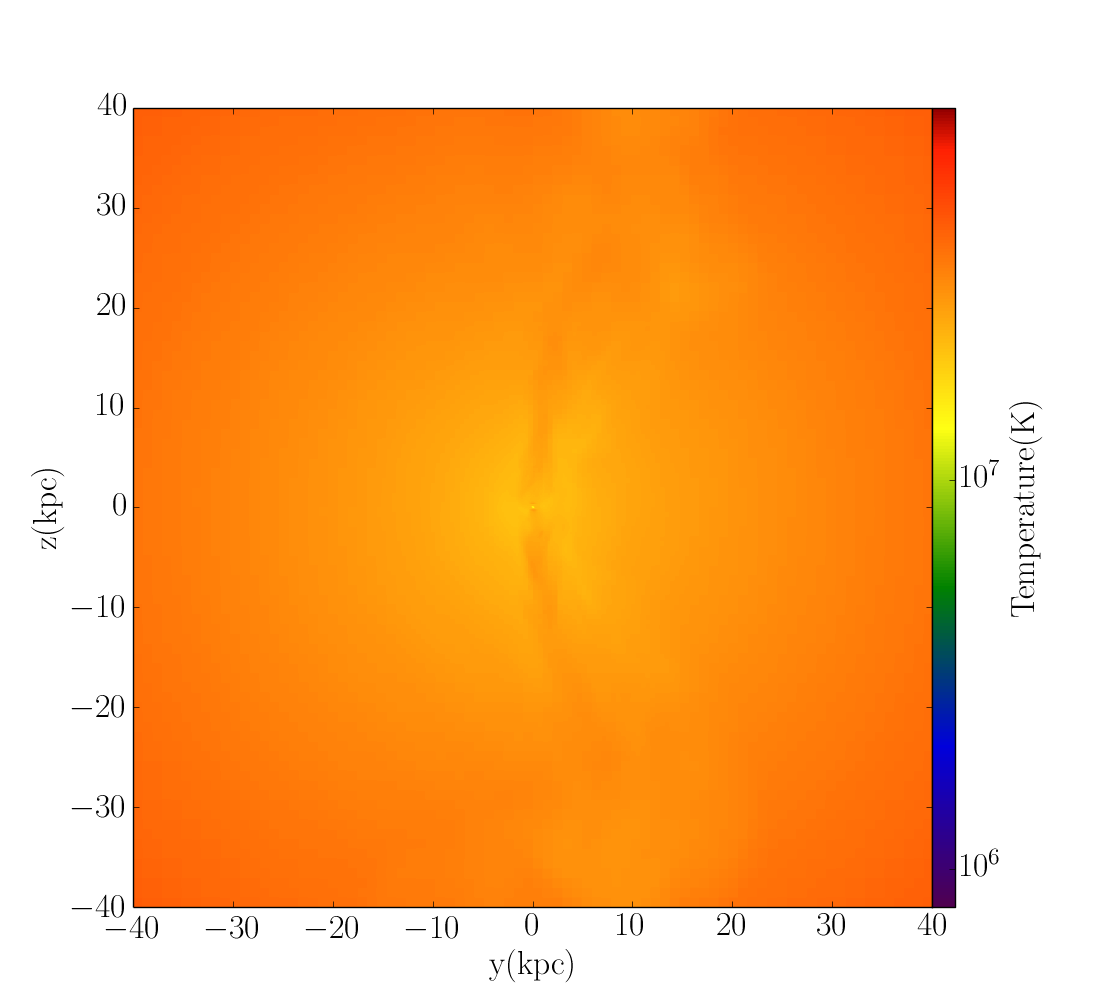}
\includegraphics[scale=.22,trim=3.2cm 2.2cm 0.6cm 0cm, clip=true]{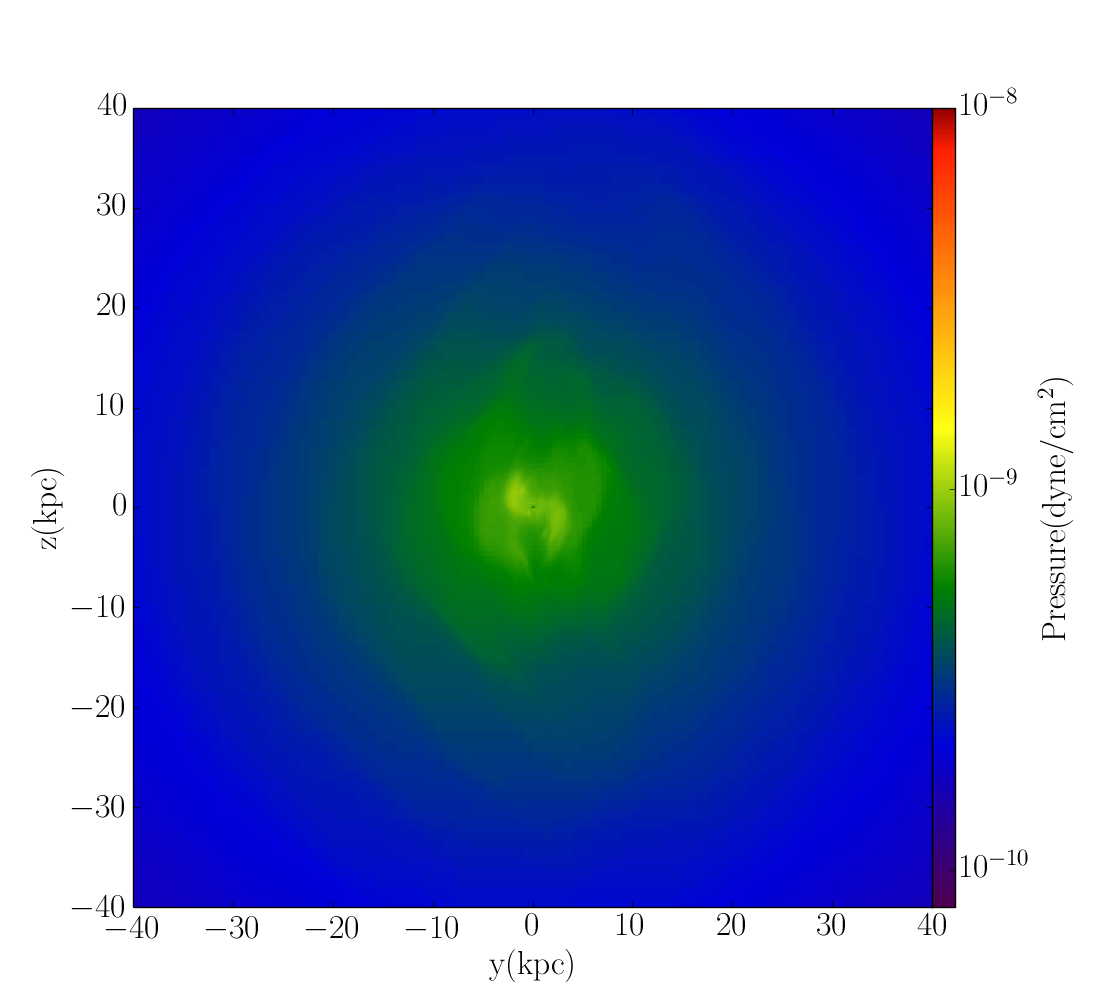}
\includegraphics[scale=.22,trim=3.3cm 2.2cm 0.8cm 0cm, clip=true]{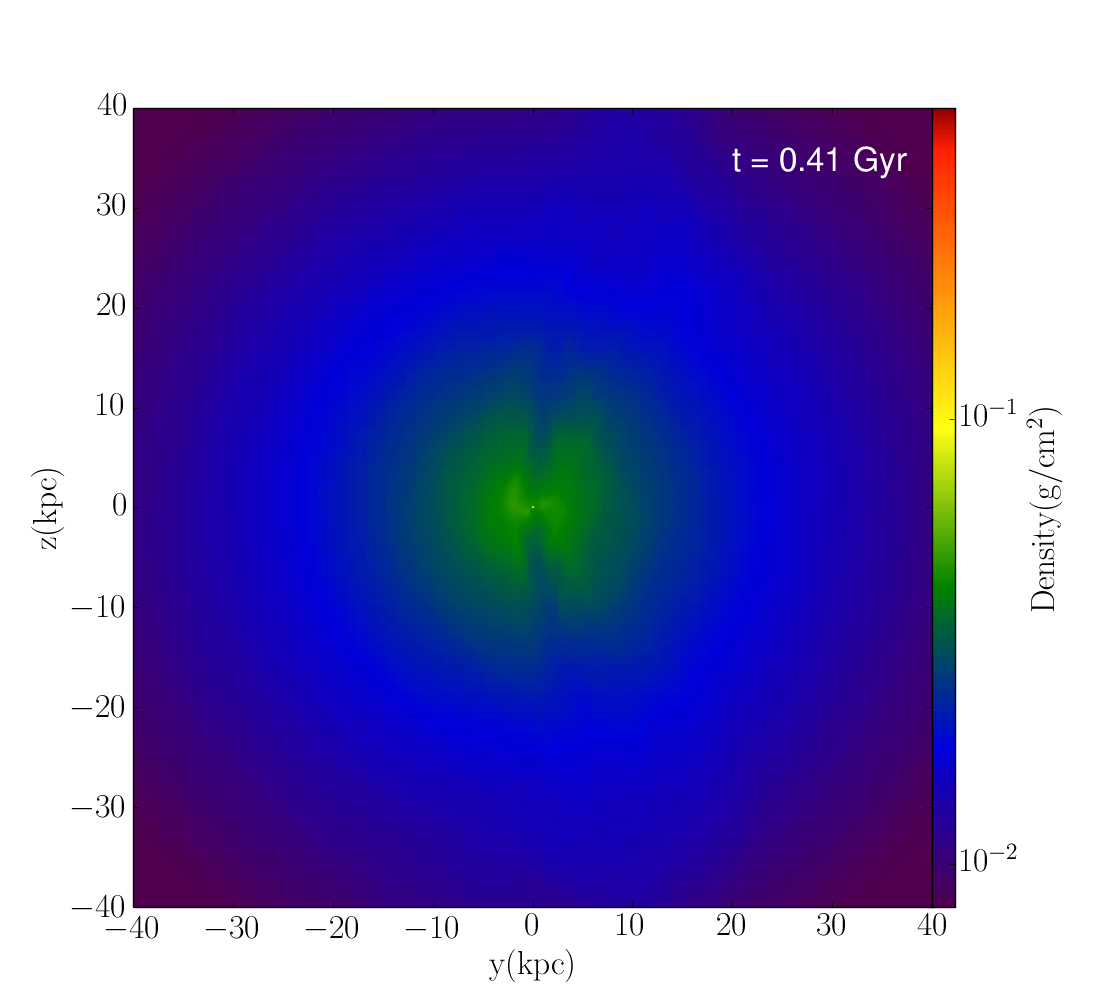}\\

\includegraphics[scale=.22,trim=0.8cm 2.2cm 0.8cm 2.3cm, clip=true]{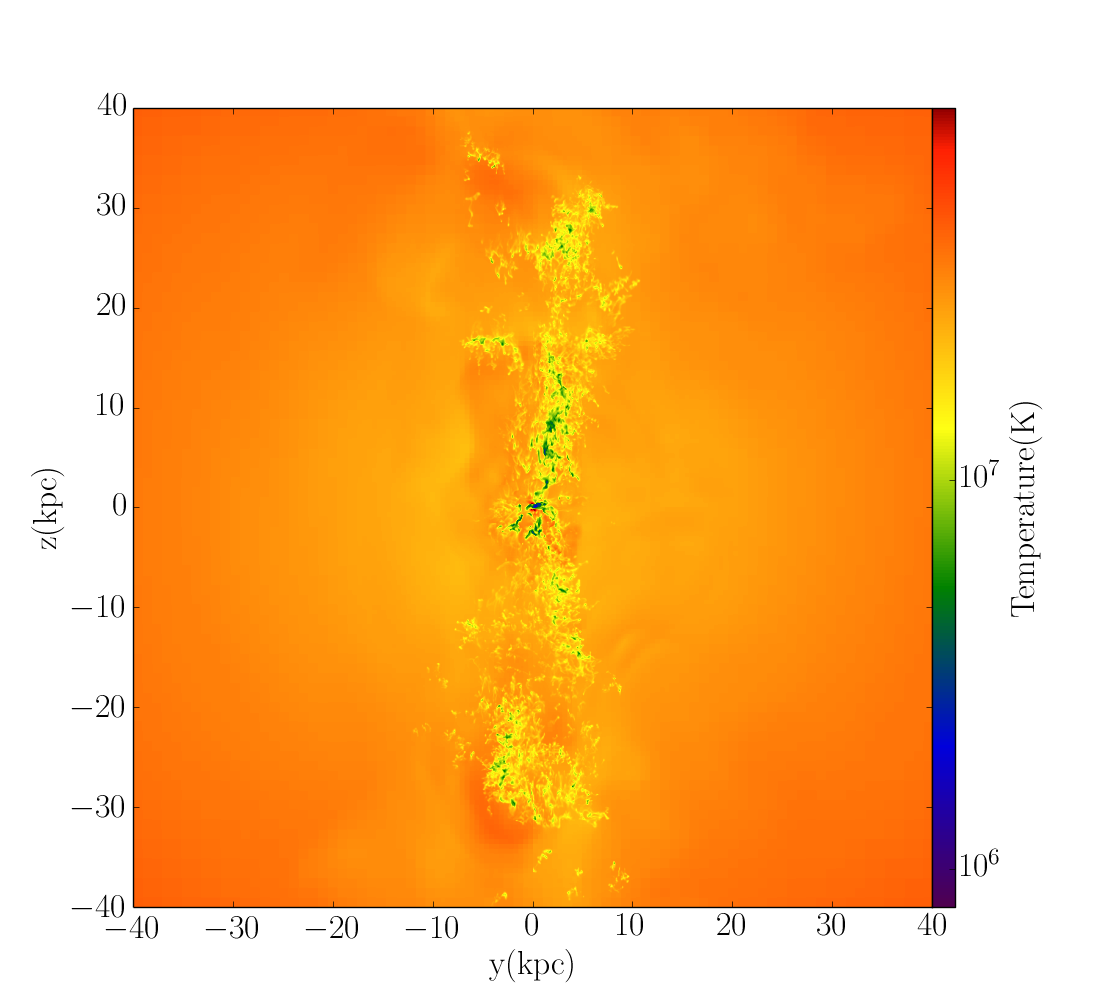}
\includegraphics[scale=.22,trim=3.2cm 2.2cm 0.6cm 2.3cm, clip=true]{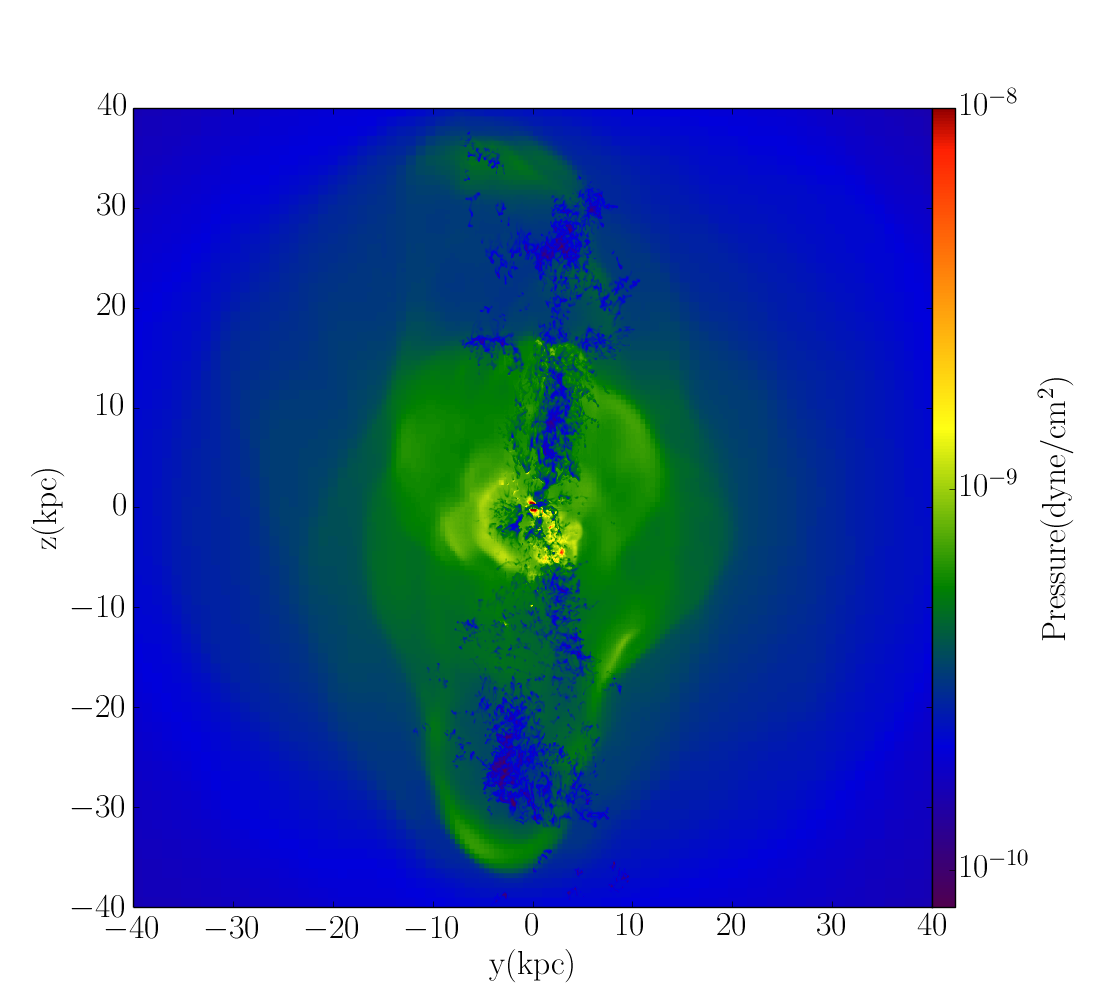}
\includegraphics[scale=.22,trim=3.3cm 2.2cm 0.8cm 2.3cm, clip=true]{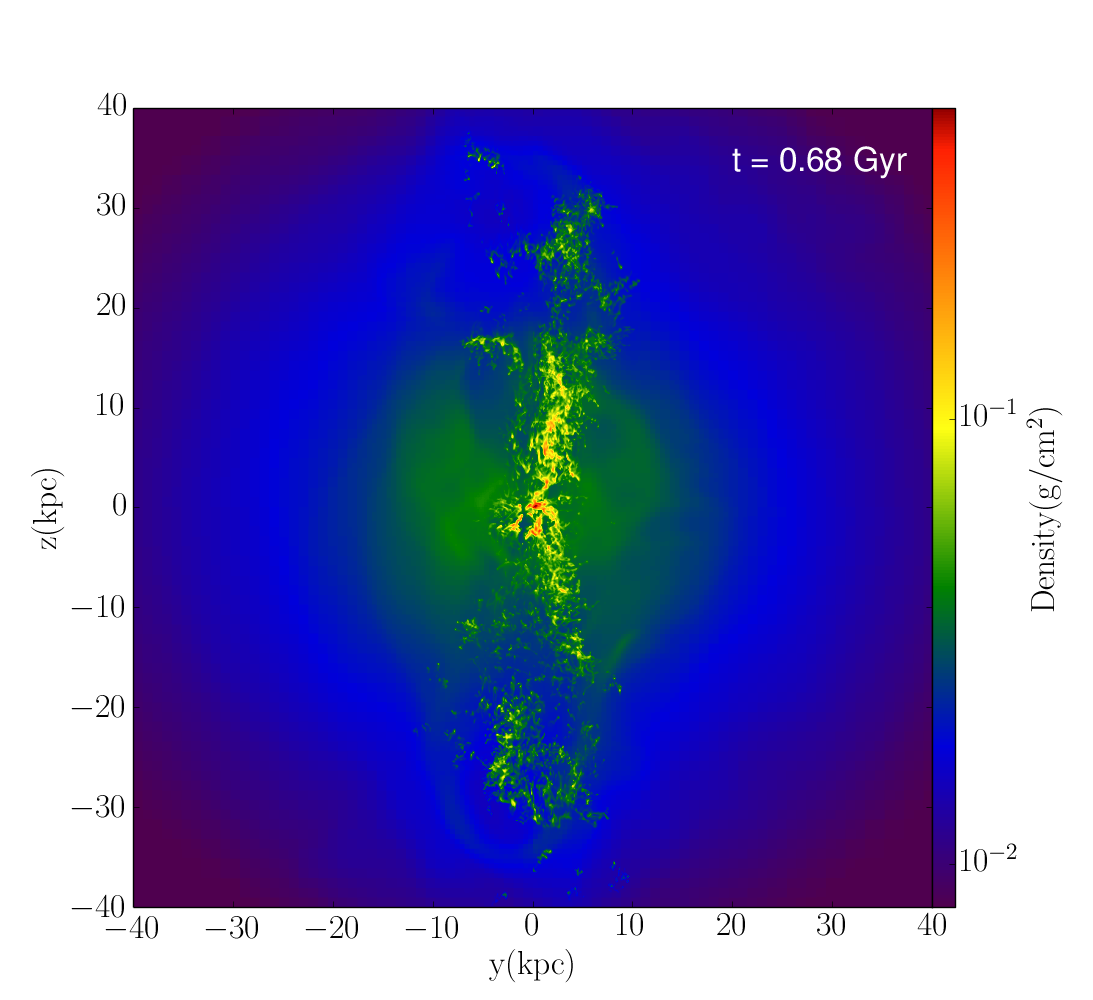}\\

\includegraphics[scale=.22,trim=0.8cm 2.2cm 0.8cm 2.3cm, clip=true]{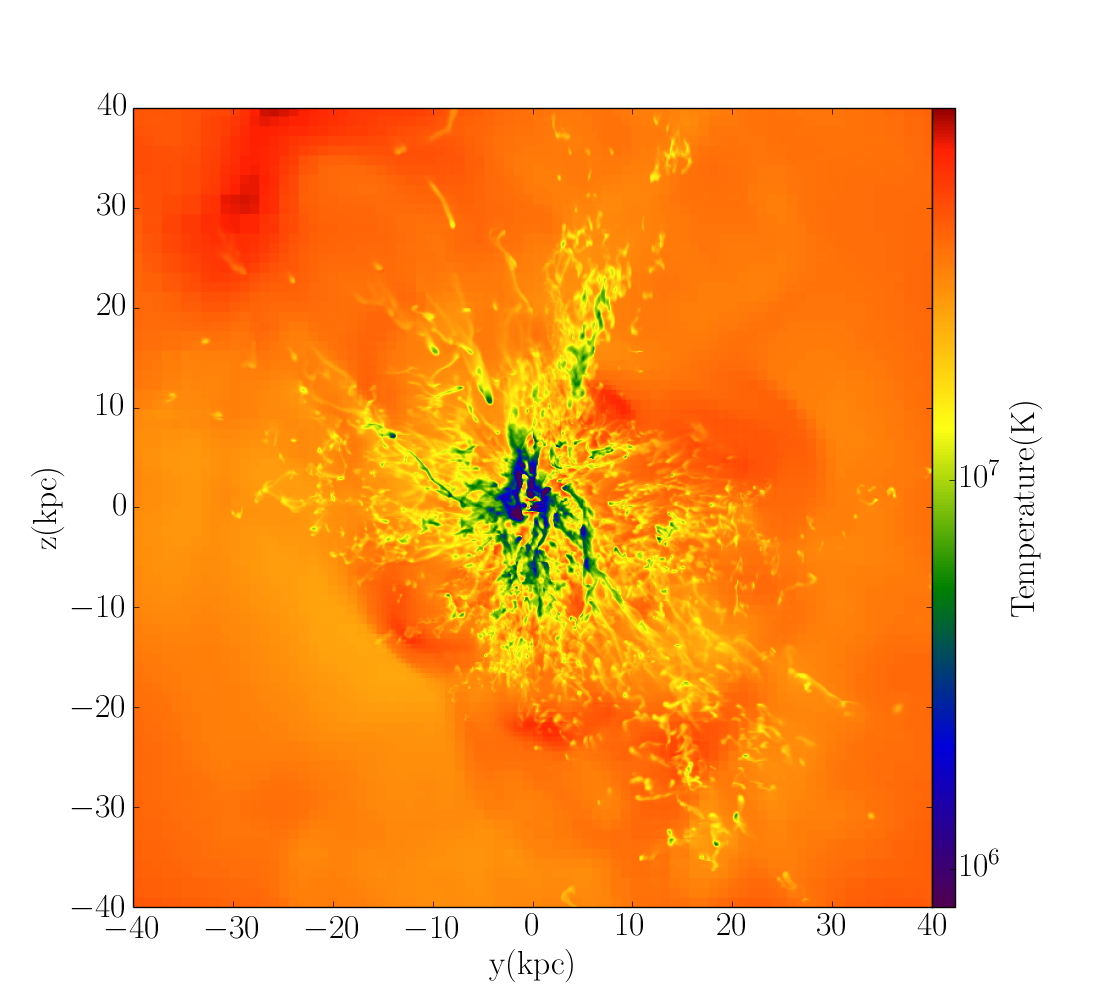}
\includegraphics[scale=.22,trim=3.2cm 2.2cm 0.6cm 2.3cm, clip=true]{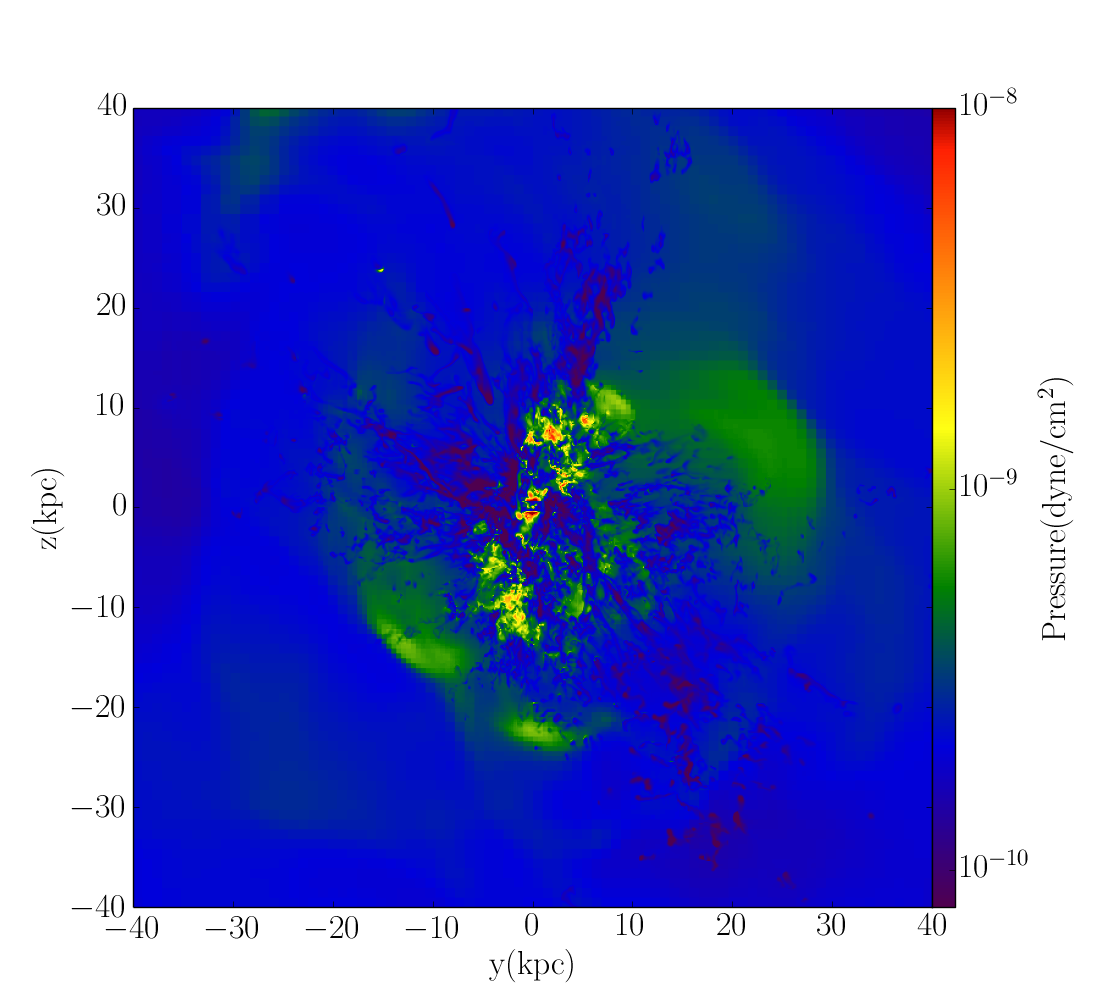}
\includegraphics[scale=.22,trim=3.3cm 2.2cm 0.8cm 2.3cm, clip=true]{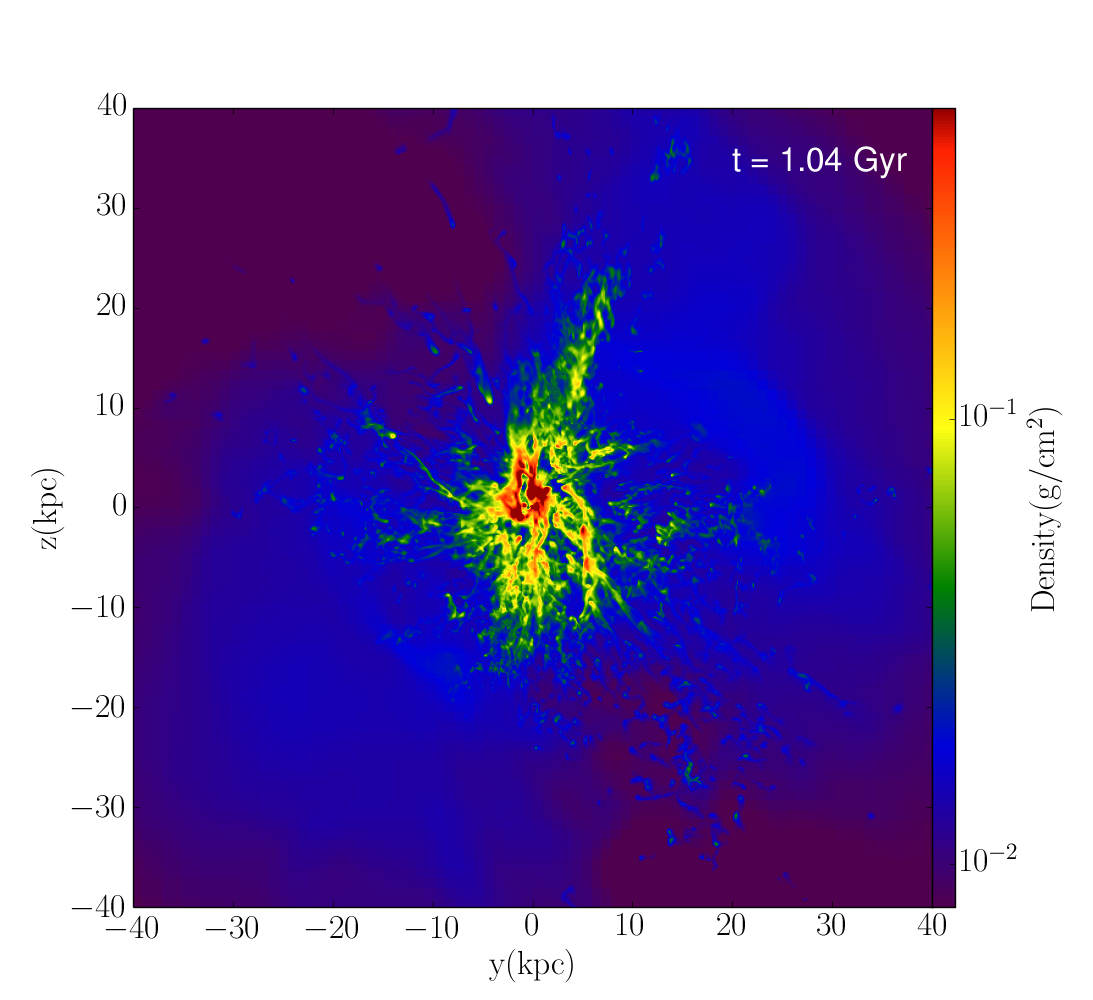}\\

\includegraphics[scale=.22,trim=0.8cm 0.5cm 0.8cm 2.3cm, clip=true]{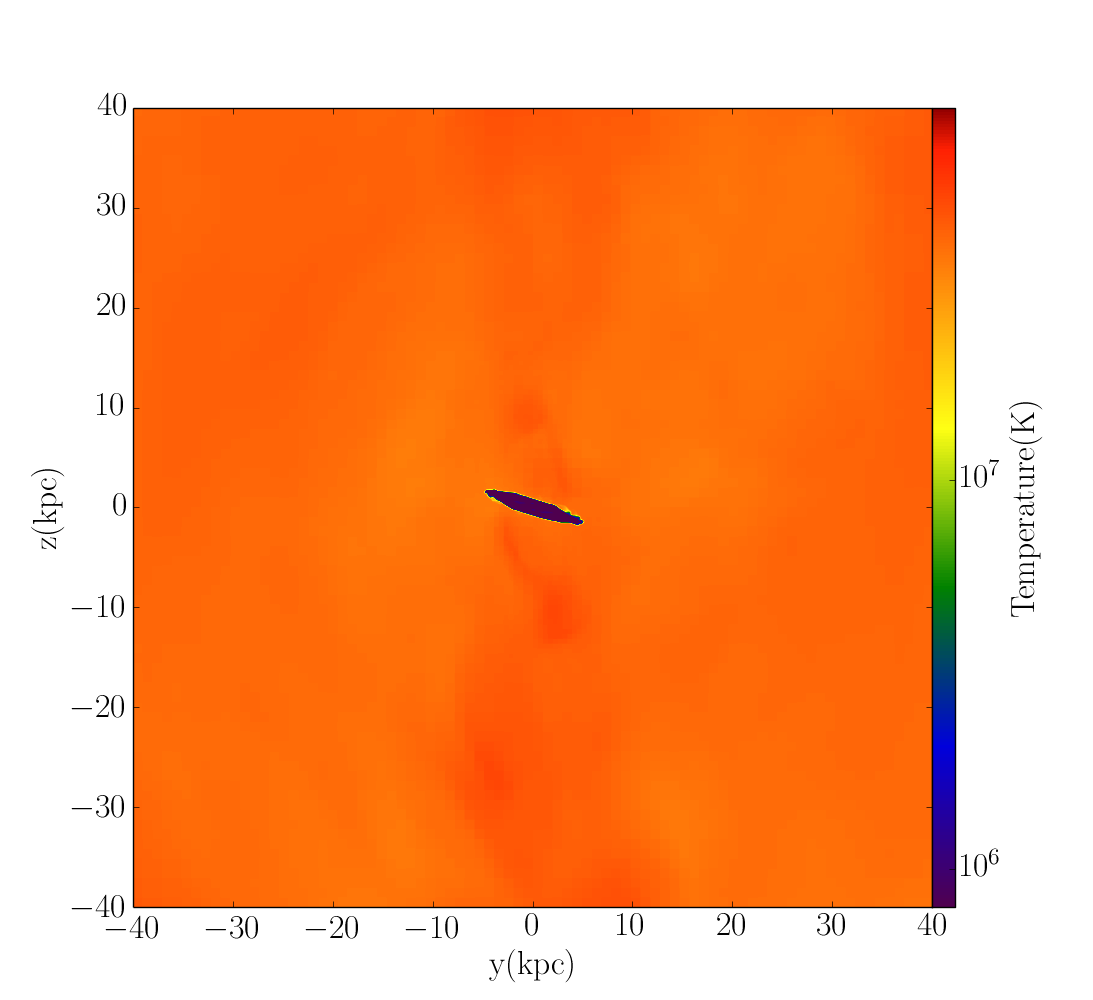}
\includegraphics[scale=.22,trim=3.2cm 0.5cm 0.6cm 2.3cm, clip=true]{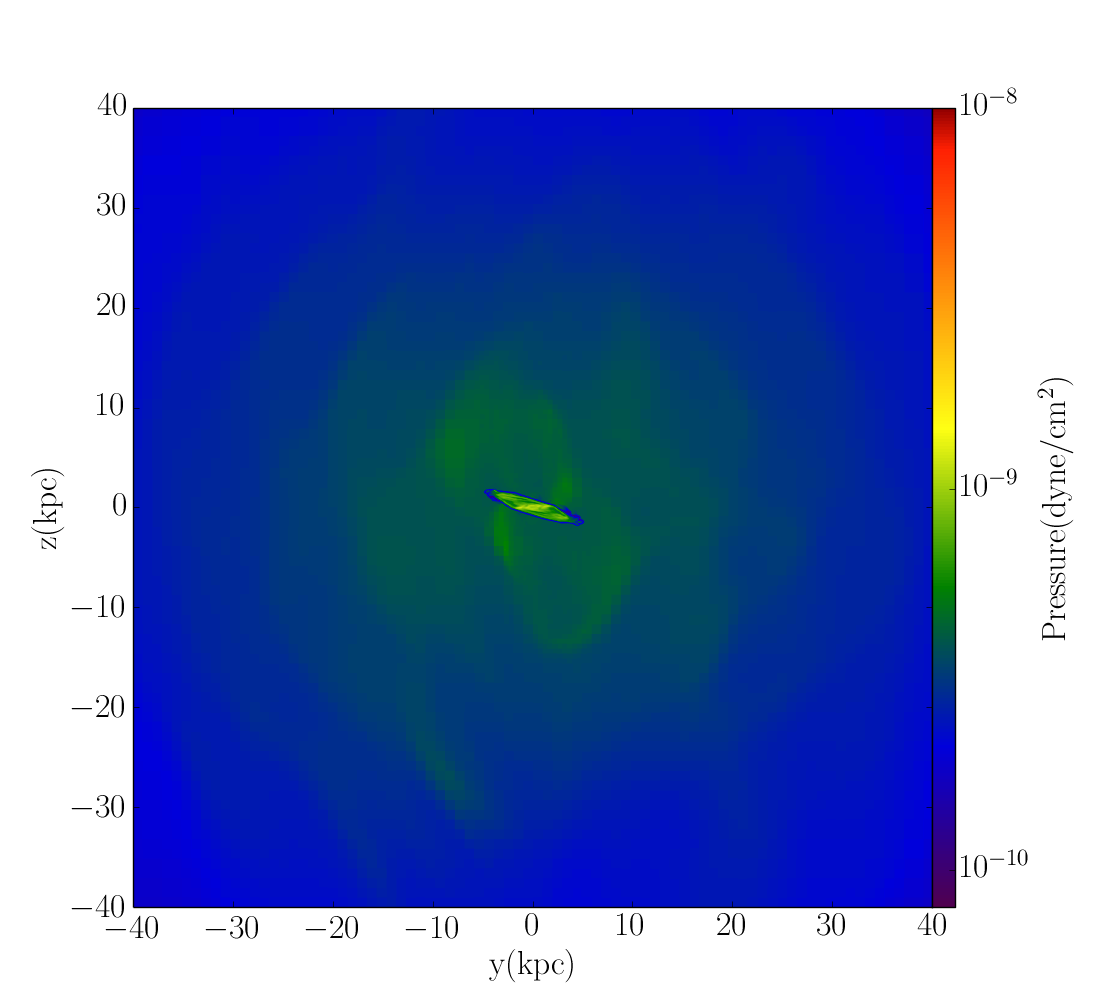}
\includegraphics[scale=.22,trim=3.3cm 0.5cm 0.8cm 2.3cm, clip=true]{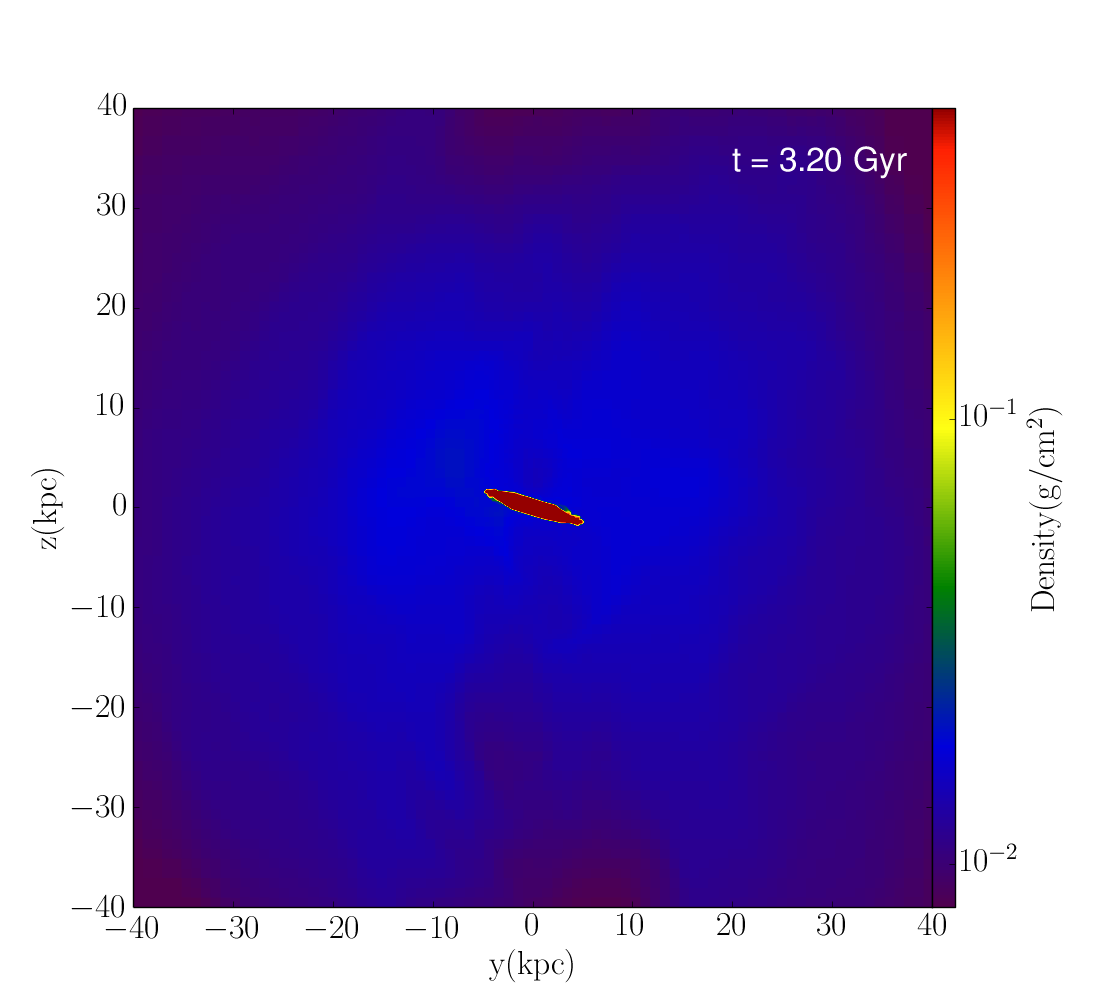}
\caption{The mass-weighted temperature (left column) and pressure (middle column), and column density (right column) of the gas in the central $r<40$ kpc region at 4 different time steps from 4 evolutionary stages projected along the x-axis. The depth of the projection box is 40 kpc. \label{fig:projections}}
\end{center}
\end{figure*}

As soon as run-away cooling happens in the center of the cluster, AGN feedback is turned on due to the presence of cold gas surrounding the SMBH. The evolution of the system can be roughly divided into four stages based on the thermal status, the characteristics of the AGN activity and the structure of the cold gas. The cluster first experiences a short period of low power AGN feedback, and then a burst of clump formation, which leads to a few Myr of chaotic cold gas condensation and accretion with a burst of AGN feedback. By $t\approx 2$ Gyr, the cluster settles to its last phase with a stable cold disk and steady AGN feedback. The details of each stage are discussed from Section~\ref{sec:results_1a} to Section~\ref{sec:results_1d}. A movie showing the time evolution of the gas temperature during the whole period (from the onset of the AGN feedback through the end of the simulation) can be found here: \url{http://vimeo.com/84807876}. The snapshots showing the gas properties in each stage are presented in Figure~\ref{fig:projections} at $t_1=0.41$, $t_2=0.68$, $t_3=1.04$ and $t_4=3.20$ Gyr.

\subsubsection{Phase One: The Onset of AGN Feedback}\label{sec:results_1a}
From $t \sim 200$ Myr to $t \sim 450$ Myr, the temperature of the gas in the central region continues to decrease and the density continues to increase, and therefore the cooling time of the gas continues to decrease. This happens in the presence of an active SMBH because the cooling rate of the gas is still low, and therefore the AGN jet power is still low, so the heating from AGN feedback is not enough to offset cooling yet. 

The first row of Figure~\ref{fig:projections} shows the projected gas density, temperature, entropy and pressure in the central $r<40$ kpc region. Like in the early pure cooling phase, gas condensation still only happens in the very center of the cluster. Outside the central cooling region, no local instability is seen to develop. 

As the cooling rate of the gas increases, the jet power steadily grows to $\sim 10^{44}$ ergs $\rm s^{-1}$ when the accretion rate increases to $\sim 2$ M$_{\odot} /$yr at the end of this stage at $t \sim 450$ Myr.

\subsubsection{Phase Two: Burst of Clump formation}\label{sec:results_1b}

Clumps of cold gas first start to form at $t \sim 450$ Myr along the propagation direction of the jets (the second row of Figure~\ref{fig:projections}). This is because the gas inside the jet propagation cone is much more perturbed than outside. A detailed analysis of the clump formation can be found in Paper II. 

Once the gas cools into clumps, it loses pressure support. The pressure of the clumps is lower than the ICM and the sound crossing time within the clumps is longer than the cooling time. Many of the newly formed clumps move outwards for a short time because of their initial positive radial velocities, but soon they rain down to the center of the cluster. Many of the clumps grow bigger along the way due to their lower pressure than the surrounding ICM, and some of them join each other to form larger filamentary structures. Some small clumps are stripped and destroyed. This process is analogous to the high velocity clouds in our galaxy \citep{Heitsch2009}.
When the clumps reach the center, due to their non-zero momentum and usually non-zero angular momentum, they oscillate and/or orbit around the SMBH while some of the cold gas gets accreted. The accretion of the cold clumps increases the jet power drastically, generating stronger shock waves, which then increases the heating rate of the ICM and thus its cooling time starts to increase, resulting in a decline of the formation of new clumps. 

\begin{figure}
\begin{center}
\includegraphics[scale=.4]{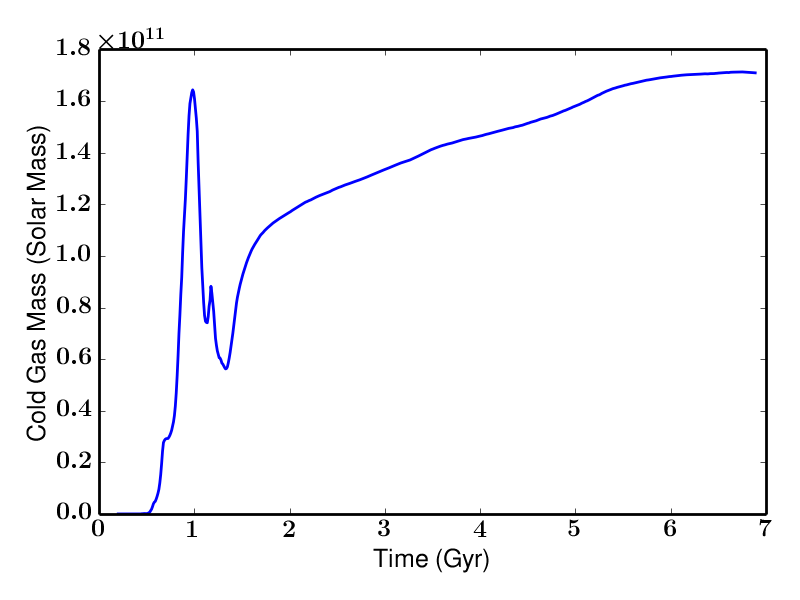}\\
\includegraphics[scale=.41,trim=0.4cm 0cm 0cm 0cm, clip=true]{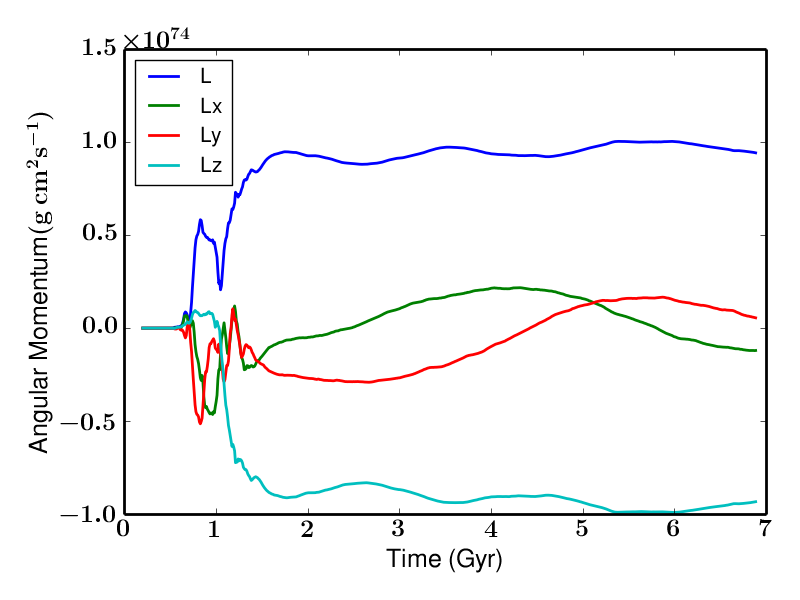}
\caption{Top: the total amount of cold gas in the central 100 kpc of the cluster as a function of time. Bottom: the angular momentum of the cold gas. The blue line shows the total angular momentum of the cold gas.
\label{fig:cold_gas}}
\end{center}
\end{figure}

This phase ends roughly when the largest cold clouds approach the center of the cluster at $t\sim 1$ Gyr. This moment is also very close to when the total amount of cold gas peaks. The evolution of the amount of cold gas in the cluster center is shown in Figure~\ref{fig:cold_gas} along with its angular momentum.

\subsubsection{Phase Three: The Co-existence of Extended Filaments and a Rotating Structure}\label{sec:results_1c}

When the largest cold clouds reach the center at $t\sim 1$ Gyr, the power of the AGN peaks. 
The cooling time of the ICM is further elevated due to the increased AGN heating, and the formation of new clumps along the z-axis is further suppressed. On the other hand, however, the gas condensation outside the original jet propagation cone (along the z-axis) starts to happen. This is related to three processes. First, the clumps that overshoot the SMBH can follow a trajectory that leads them off the z-axis due to their non-zero angular momentum. As they travel through the region to the side of the jet, they can both grow larger themselves, and perturb the local ICM along the way. Second, the jet has been on for a few hundred Myr by this time and the power of it has increased, both of which help establish a global turbulence on a larger scale throughout the core. Last but most importantly, when large clouds of cold gas are present around the SMBH, if they are in the way of the jet, instead of being destroyed as small clumps might be, they force the jet out through lower density regions, which effectively changes the direction of the jet. This physical process is similar to the quasar outflow confined by the cold disk in a galaxy \citep{Claude2012}. 
The motion of the cold gas surrounding the SMBH is very chaotic, so the jet comes out in all random directions, perturbing the ICM everywhere in the cluster core. The third row of Figure~\ref{fig:projections} shows that at $t_3=1.04$ Gyr, the jets come out at about $45^{\circ}$ from the z-axis, causing clumps to form in that direction. The extended cold gas at this time looks strikingly similar to the $\rm H\alpha$ structure in the Perseus cluster \citep{Conselice2001}. The chaotic accretion of cold clouds is also seen in \citet{Gaspari2013} with cooling and continuous injection of turbulence.

Like the cold clumps that cool along the z-axis earlier, the clumps formed in other directions also fall towards the center of the cluster almost ballistically. The cold clouds that have already accumulated in the center collide and combine into a rotating structure. While gas in the central area of the rotating gas cloud is accreting onto the SMBH, more clumps fall down to join the cloud. Because of the abundance of cold gas surrounding the SMBH, even though the cooling rate of the ICM has been reduced due to the AGN outburst, the jet power is still high and is only slowly declining. The heating rate is still higher than the ICM cooling rate. This further elevates the temperature and the cooling time of the ICM, making clump formation more and more difficult.

After another few hundred Myr, at $t\sim 1.7$ Gyr, clump formation becomes very rare and all the cold gas has settled to a rotating disk: the cluster enters its last phase.

\subsubsection{Phase Four: A Long-lasting Massive Cold Disk}\label{sec:results_1d}

The last stage lasts for a few Gyr through the end of our simulation. The physical properties of the ICM stay rather constant, indicating that a thermal balance is achieved and maintained over a few Gyr. 

The ICM in the core of the cluster is still cooling, but it cools in a fashion more similar to the first stage, i.e., a reduced cooling flow is present, but gas only condenses out of the flow in the center of the cluster instead of through local instabilities as in the previous two stages. Clump formation at $r>10$ kpc becomes extremely rare. In the center of the cluster, the reduced cooling flow (of hot ICM) cools directly onto the cold disk through a mixing layer surrounding the disk. The cold disk has a size of $r \approx 6$ kpc and a total mass of $\sim 10^{11}$ M$_{\odot}$. Note that the cold disk is hollow in the very center (the accretion zone), but due to its thinness, we refer to it as a ``disk'' for simplicity instead of calling it a thin torus (see the bottom row of Figure ~\ref{fig:projections}).

\begin{figure}
\begin{center}
\includegraphics[scale=.4]{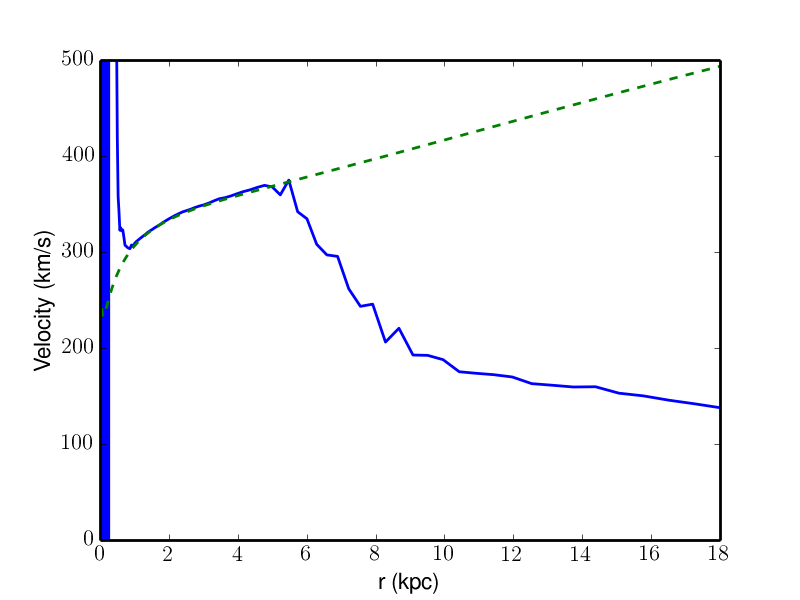}
\caption{The rotational velocity (blue line) of the gas compared with the Keplerian velocity (green dashed line) at $t_4=3.18$ Gyr. Due to the possible warp of the cold disk, we calculate the cell mass-weighted tangential velocity as a proxy for the rotational velocity of the gas. The rotational velocity follows the Keplerian velocity within the disk ($r \lesssim 5$ kpc), indicating that the disk is rotationally supported.
\label{fig:kepler}}
\end{center}
\end{figure}

The disk is rotationally supported as shown in Figure~\ref{fig:kepler}. When the hot ICM cools onto the disk, it adds mass but not enough angular momentum to the disk because the hot ICM is not rotationally supported. As is shown in Figure~\ref{fig:kepler}, the rotational velocity of the gas deviates from the Keplerian velocity outside the disk. The disk self-adjusts such that the low angular momentum material is lost in the center, which then gets accreted to the SMBH and the bulk of the disk stays rotationally supported. Therefore, the AGN jet power, which is proportional to the accretion rate, is still closely (although not directly) linked to the ICM cooling rate through the disk. 

The mass of the disk grows by about $50\% $ from $t \sim 2$ Gyr to $t \sim 7$ Gyr at the end of the simulation (see Figure~\ref{fig:cold_gas}), which corresponds to an average growth rate of about 10 M$_{\odot}/\rm yr$. The initial rotation of the disk is determined by the total angular momentum of the cold clumps formed at large distances in early stages, and is therefore random. In the long term, the ICM does add angular momentum to the disk, which is not necessarily aligned with the initial angular momentum of the disk. The interaction between the jet material and the disk can also cause a secular evolution in its rotation. Therefore, the rotation direction of the disk changes slowly over a few Gyr timescale, which is shown in the movie and the right panel of Figure~\ref{fig:cold_gas}.

\subsection{Balance between Heating and Cooling}\label{sec:results_heating_cooling}

In this section, we analyze how AGN heating balances ICM cooling in the simulation. 

We first examine heating under the microscope. The shock waves observed in nearby cool core clusters are generally rather weak with only a few exceptions \citep[see review by][]{McNamara2007}. To measure the strength of the shocks in the simulation, we randomly sample lines drawn from the center of the cluster outwards. We measure the gas properties along the lines and identify the location of shocks. Then the Mach numbers of the shocks are estimated based on the pressure jump: 
\begin{equation}
\frac{P_2}{P_1}=1+\frac{4\gamma}{\gamma +1}(M-1) \; .
\end{equation}

\begin{figure*}
\begin{center}
\includegraphics[scale=.44]{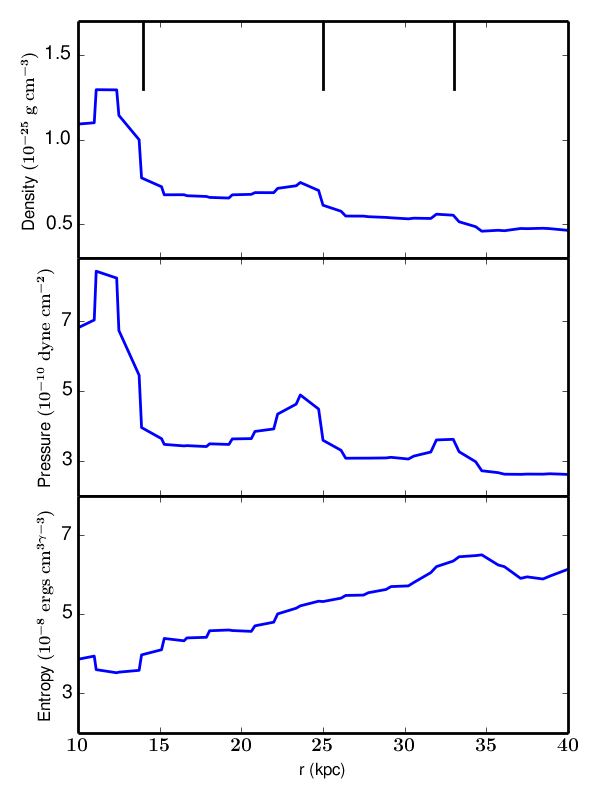}
\includegraphics[scale=.38]{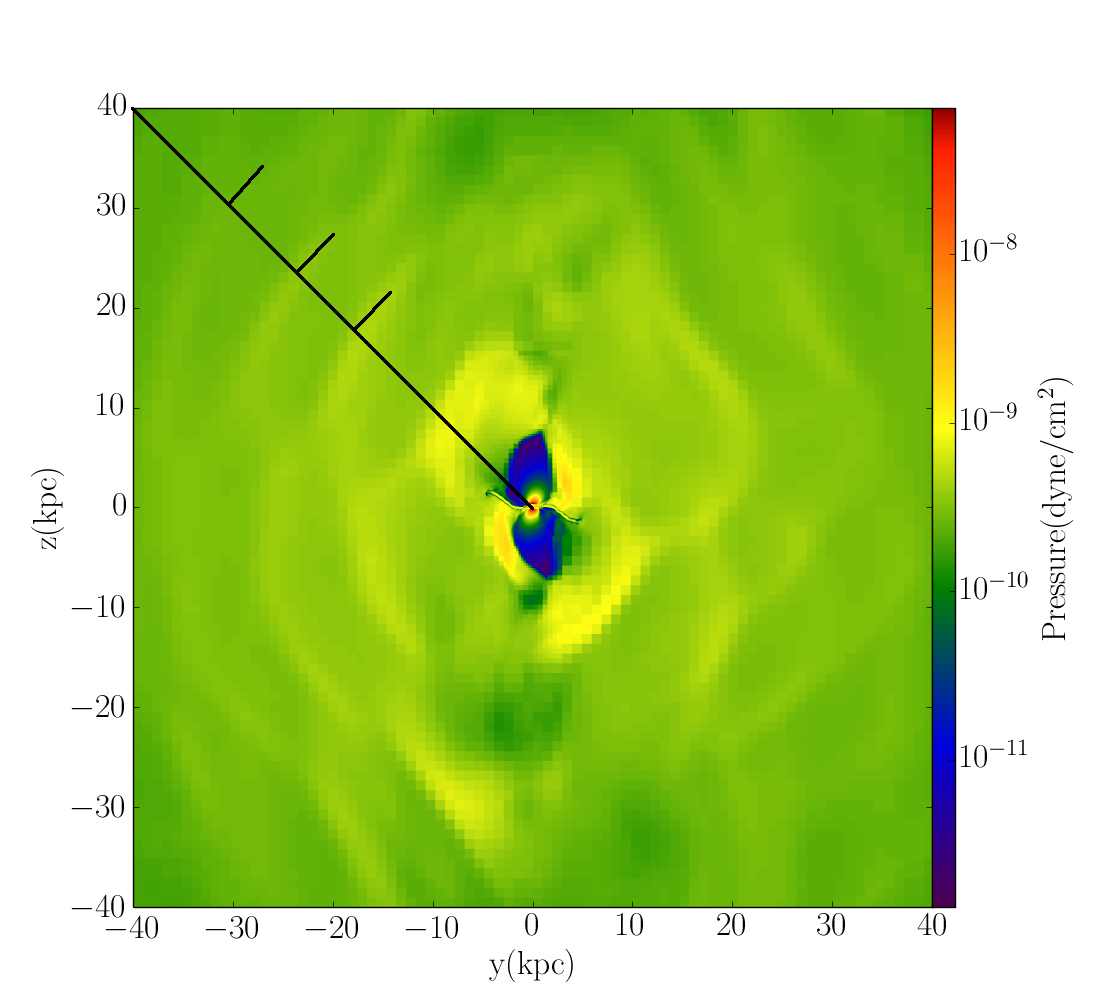}
\caption{Left: the gas density, pressure and entropy along a line drawn from the center of the cluster (shown as a black solid line on the right panel). Right: the pressure map of a slice of gas through the center of the cluster in the y-z plane at $t_4=3.2$ Gyr. The three short black lines indicate where the Mach numbers of the shocks are measured in both panels. 
\label{fig:mach}}
\end{center}
\end{figure*}

Most of the shockwaves at $r>5$ kpc are weak shocks with a Mach number only slightly above 1. As an example, Figure~\ref{fig:mach} shows the pressure map of a slice of gas at $t_4=3.2$ Gyr and the gas properties along a line. The entropy is roughly the same across the shock front, and the Mach numbers measured at $r=$14, 25 and 33 kpc are 1.15, 1.13, and 1.04, respectively. These shocks are consistent with the weak shocks observed in nearby cool core clusters \citep[e.g.][]{Fabian2006, Blanton2009}.

The shock heating time scale can be roughly estimated as 
\begin{equation}
t_{sh} = \frac{E_{thermal}}{f_{sh}\Delta E_{thermal}}\; ,
\end{equation}
where $E_{thermal}$ and $\Delta E_{thermal}$ are the thermal energy and its average increase after each shock, and $f_{sh}$ is the shock passing frequency. For weak shocks, 
\begin{equation}
\frac{\Delta E_{thermal}}{E_{thermal}}=\frac{\Delta T}{T} \approx M-1\; .
\end{equation}
Given the typical separation between two successive shocks of $d\sim 10$ kpc (Figure~\ref{fig:mach}) and the sound speed $Cs\sim 800 \ \rm{km/s}$ at a few 10s kpc, 
\begin{equation}
t_{sh}=\frac{d}{Cs} \frac{1}{M(M-1)} \approx \frac{1}{M(M-1)}\times 10^7 yr.
\end{equation}
For $M=1.05-1.1$, the shock heating timescale $t_{sh} \approx 100 - 200$ Myr. This is comparable to the cooling time of the gas at a few tens kpc. Although shock heating is important, other physical processes including sound wave \citep{Mathews} and turbulent mixing \citep{Kim2003} also contribute to the heating of the ICM.

\begin{figure}
\begin{center}
\includegraphics[scale=.4]{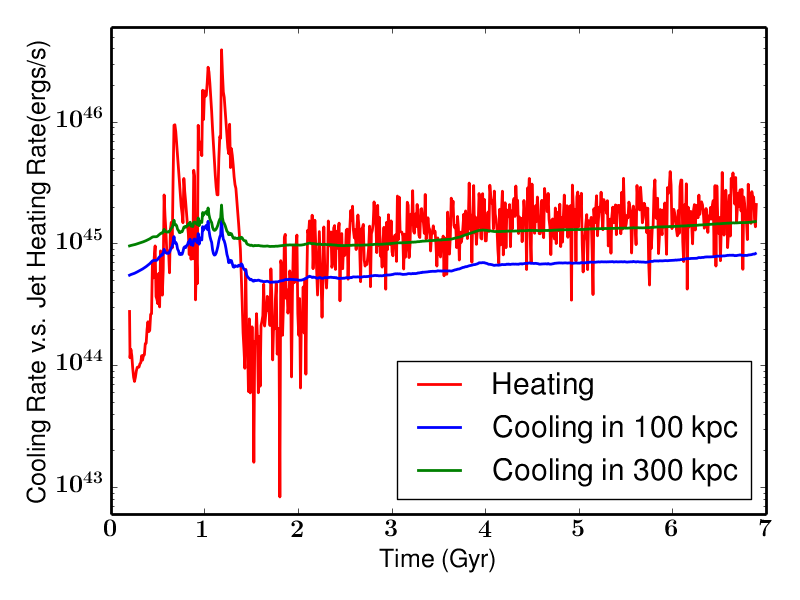}\\
\includegraphics[scale=.4]{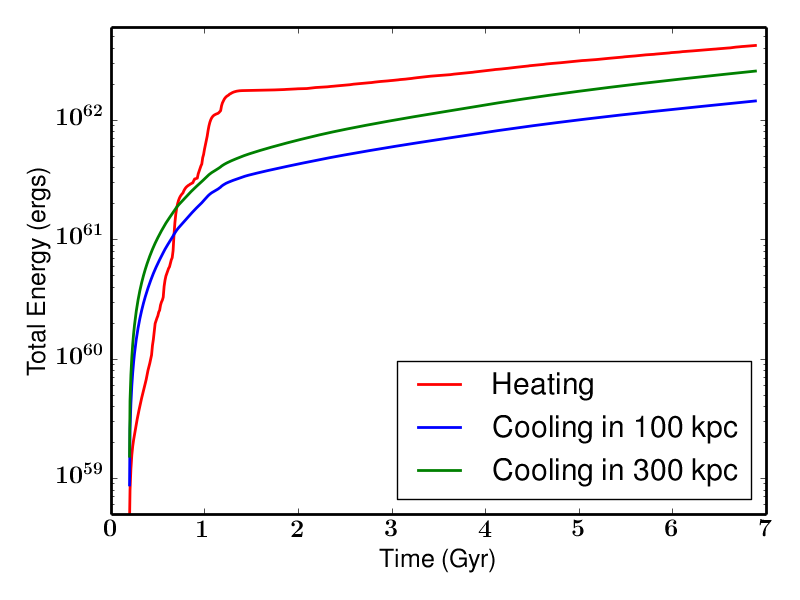}
\caption{Top: the jet heating rate ($\dot{E}$) and the total cooling rate in the central $r<100$ kpc and 300 kpc region of the cluster as a function of time. Bottom: the total energy output from the SMBH (the cumulated $\dot{E}$) and the total energy loss due to radiative cooling in the central 100 kpc and 300 kpc of the cluster as a function of time.
\label{fig:heating}}
\end{center}
\end{figure}

Over a few Gyr timescale throughout our simulation, the ICM cooling is well balanced by the AGN heating. As is shown in the left panel of Figure~\ref{fig:heating}, the instantaneous jet power closely follows the total cooling rate in the core of the cluster, calculated as the total energy loss rate through radiation within $r<100$ kpc and $r < 300$ kpc. Noticeably, three large peaks are seen in all three curves at $t \sim$ 0.7, 1.0, and 1.2 Gyr. The cooling rate peaks when a large amount of ICM cools into clumps. There is a short delay of a few tens of Myr between the peak of the jet heating and the peak of the cooling. This is the time it takes for the cold clumps to fall to the SMBH in the cluster center, which is of the order of magnitude of the dynamical time of the gas at a few 10s kpc where the clumps form. There is another delay between the peak of the AGN power and when the ICM is heated up, shown as the local minima of the cooling rate. This delay again is very short as shown in Figure~\ref{fig:heating}. A rough estimation gives $\rm \frac{100 \ kpc}{1000 \ km/s} \approx 100$ Myr, where 100 kpc is the core size and 1000 km/s is the typical shock velocity (close to the sound speed for weak shocks).

The total energy output from the AGN exceeds the cooling loss by $t\sim 0.7$ Gyr (right panel of Figure~\ref{fig:heating}), and starts to grow steadily with roughly the same slope as the energy loss due to radiative cooling once the disk stabilizes at $t\sim 1.7$ Gyr. The heating is efficient though not perfect. Less than $30\%$ of the energy goes to balance cooling within 100 kpc, and at most $60\%$ is deposited within 300 kpc. The rest leaks out of the core.

\begin{figure}
\begin{center}
\includegraphics[scale=.45]{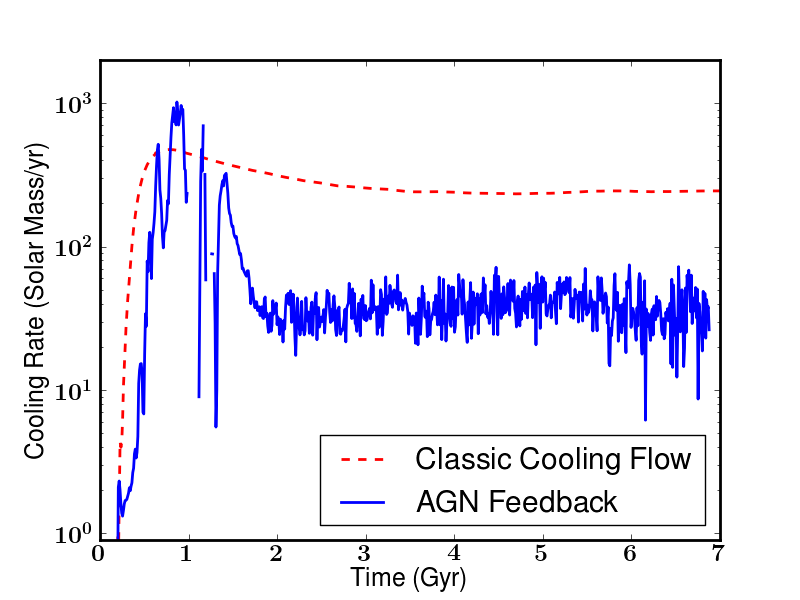}
\caption{The mass cooling rate, i.e. the rate at which the ICM cools into cold gas, shown as the blue line. The red dashed line shows the cooling rate in our pure cooling flow simulation without the AGN feedback, which gives a classic cooling flow rate of about $\sim 250 \rm M_{odor}/yr$. Our standard run with AGN feedback suppresses cooling by roughly an order of magnitude. 
\label{fig:Mdot}}
\end{center}
\end{figure}

The cooling rate can also be expressed as the rate at which the ICM cools into cold gas, also referred to as the ``mass cooling rate'' or the ``mass deposition rate''. In our simulations, this is the growth rate of the total amount of cold gas plus the SMBH accretion rate. Figure~\ref{fig:Mdot} shows how this mass cooling rate in our standard run compares with our pure cooling flow simulation with the same initial condition and resolution but without the AGN. With AGN feedback, the cooling rate only reaches the classic cooling flow value very briefly at the peak of the clump formation stage. On average, the cooling rate in our standard run is about 30 $\rm M_{\odot}/$yr, roughly an order of magnitude lower than the classic cooling flow rate of $\sim 270 \ \rm M_{\odot}/$yr \citep{Allen1992}.

\begin{figure*}
\begin{center}
\includegraphics[scale=.4]{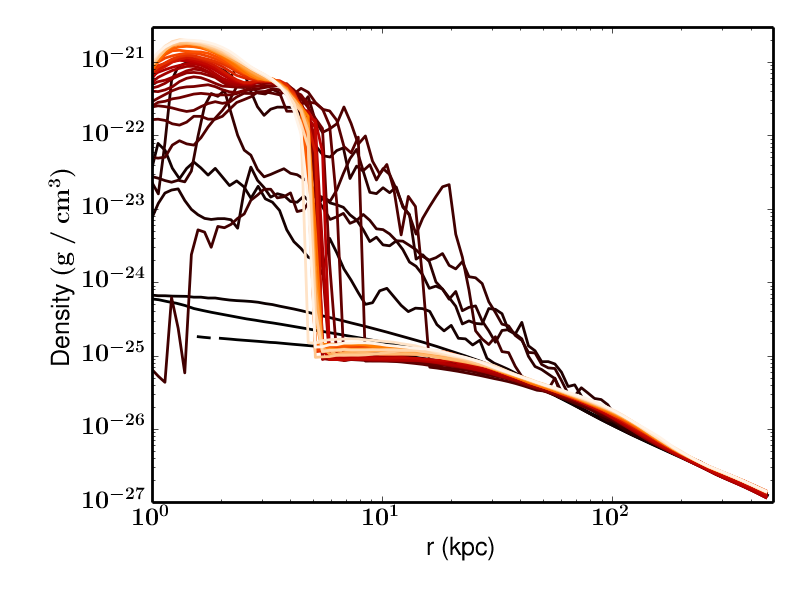}
\includegraphics[scale=.4]{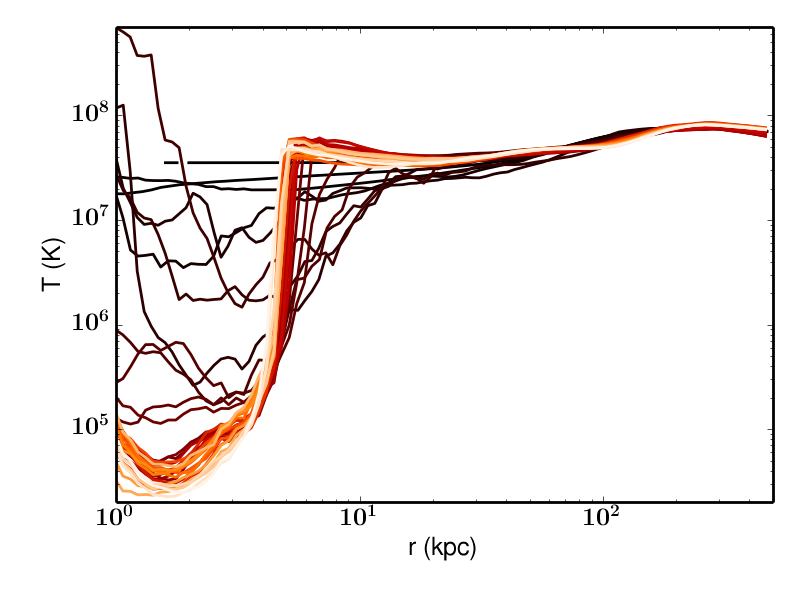}\\
\includegraphics[scale=.4]{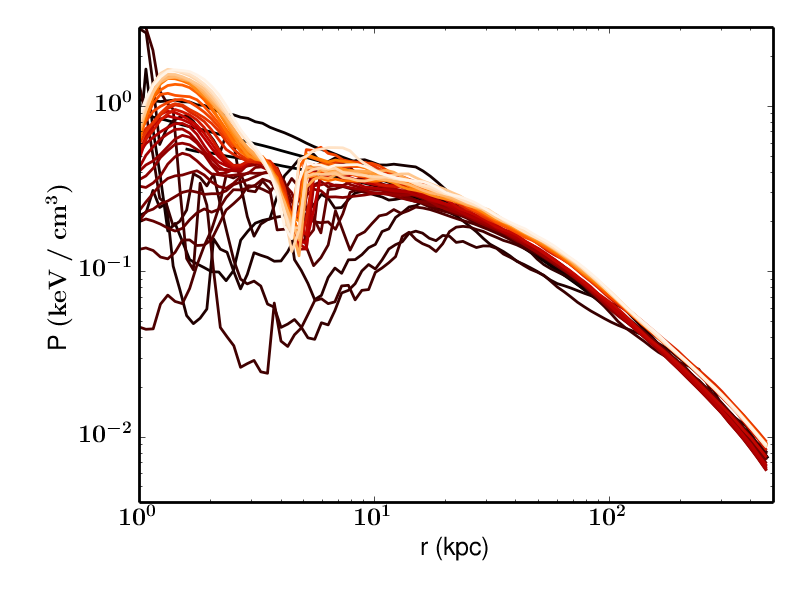}
\includegraphics[scale=.4]{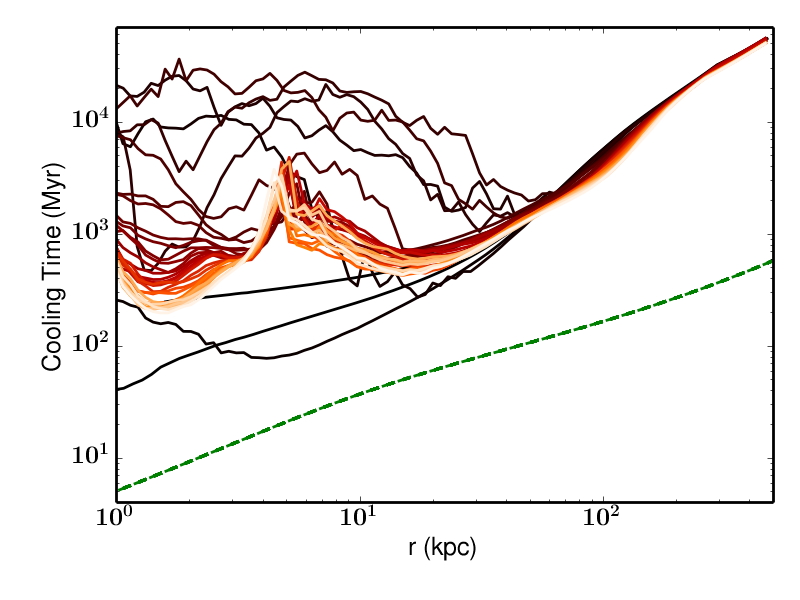}
\caption{The radial profiles of gas density, temperature, pressure and cooling time weighted by cell mass at different times (from dark brown to light yellow, the darkest being the initial condition). The time interval between two successive lines is 20 Myr. The green dashed line on the last panel shows the dynamical time of the gas. 
\label{fig:profile}}
\end{center}
\end{figure*}

Figure~\ref{fig:profile} shows the radial profiles of the cell mass-weighted gas density, temperature, pressure and cooling time throughout the simulation. The profiles stay unchanged at $r > 200$ kpc due to the long cooling time of the hot gas at large radii. At $r \sim 100$ kpc, the gas cools slowly and rather smoothly over the course of the simulation, but the change is very small. Inside a few tens kpc, the gas first becomes cooler and denser before the jet power builds up. The cooling time is the shortest right before clumps start to form. As a reference, the dynamical time is shown as the green line. The profiles show large fluctuations during the clump formation phase due to the existence of cold clumps and the high jet power. The temperature is slightly elevated. During the last phase, the profiles stay rather constant over a few Gyr timescale and a thermal balance is maintained as is discussed in Section~\ref{sec:results_1d}. The temperature gradient stays positive at $\sim 100$ kpc, indicating that a cool-core appearance of the cluster is also maintained.

\subsection{Simulation Comparison}\label{sec:results_other_simulations}
Besides the standard run discussed above, we also perform a set of test runs to study the impact of resolution and the choice of parameters in our model. In this section, we discuss the main results of each test.

\subsubsection{Resolution}\label{sec:results_resolution}
To test the dependence of our results on resolution, we perform simulations with maximum refinement level $l_{max}$ of 11, 10, 9 and 8, corresponding to the physically size of the smallest cell $\Delta x_{min}$ to be $\sim 0.12, 0.24$, 0.5 and 1 kpc. All the simulation parameters are the same except that we double the physical size of the accretion region (to be 2 kpc) for the two lowest resolution runs and reduce $r_{jet}$ and $h_{jet}$ (which still give larger values in physical units than the standard run) as described in Section~\ref{sec:methodology_jet}.

The results of the simulations with $l_{max} = 10$ and 11 (hereafter refereed to as ``the medium resolution simulations'') are very similar to our standard simulation, indicating convergence at resolutions better than a few hundred kpc. The total number of clumps formed at early times decreases with lower resolution, as we discussed in Paper II, but the overall cluster evolution stays the same for all these runs.

As we have discussed in Section~\ref{sec:results_1d}, the initial angular momentum of the cold gas is random. This is confirmed in the medium resolution simulations: the run with $l_{max} = 10$ has a disk that is initially almost aligned with the y-z plane. 

The simulation with $l_{max} = 9$ (hereafter refereed to as ``the low resolution simulation'') however, shows a different initial evolution. The clumps first form around the same time as the standard run, but both the number of clumps and the amount of gas that condenses into clumps are lower. We have shown in Paper II that the ability to resolve clumps is limited by resolution. In addition, the physical size of the jet launching planes is larger in lower resolution simulations, and therefore, the jet material has a lower density given $\dot{M}$, which results in a weaker perturbation, and a slower buildup of turbulence. Cold clumps are also more vulnerable to destruction with lower resolution. The cold gas falls to the center and forms a rotating disk. However, unlike in the standard run, the disk is quickly accreted onto the SMBH without any cold gas left, because the disk is smaller and the angular momentum is less well conserved here due to low resolution. The AGN is turned off and the ICM starts to cool again. At $t\sim 2$ Gyr, another cycle begins with cooling catastrophe happening in the very center of the cluster, triggering the AGN feedback which leads to another burst of clump formation. The second cycle ends at $t\sim 3$ Gyr with the complete accretion of the cold gas. The AGN is turned off again and the cluster enters a quiescent pure cooling phase which lasts for a few hundred Myr. The cluster experiences a third cycle which ends  at $t \sim 5.5$ Gyr. 

By the time the fourth cycle starts at $t\sim 6$ Gyr, the gas temperature has decreased by $1-2$ keV at $r\sim 80 - 200$ kpc, where the cooling time has decreased. In the fourth cycle, when the disk forms, cooling overwhelms heating and the growth of the disk is faster than its consumption by the SMBH. The disk stays and grows to $\sim 4 \times 10^{11}$ M$_{\odot}$ at the end of the simulation at $t=15$ Gyr.

\begin{figure}
\begin{center}
\includegraphics[scale=.4]{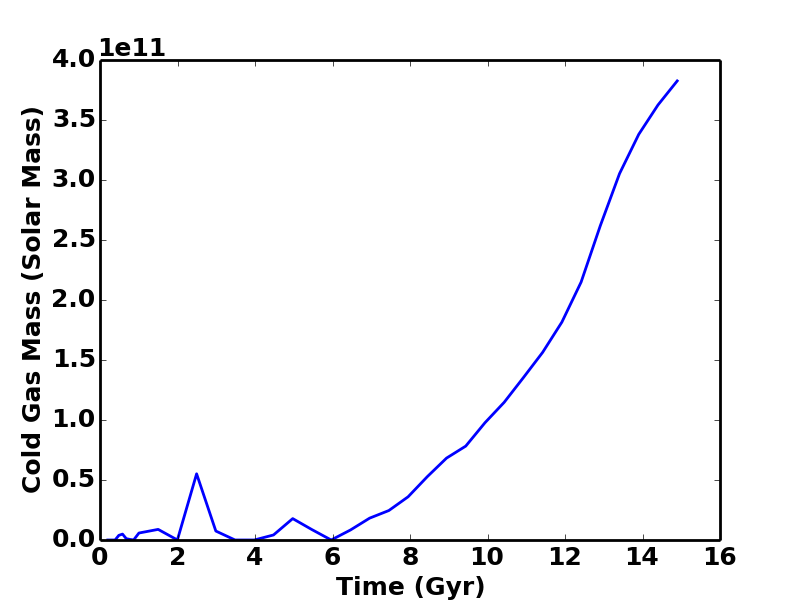}\\
\includegraphics[scale=.4]{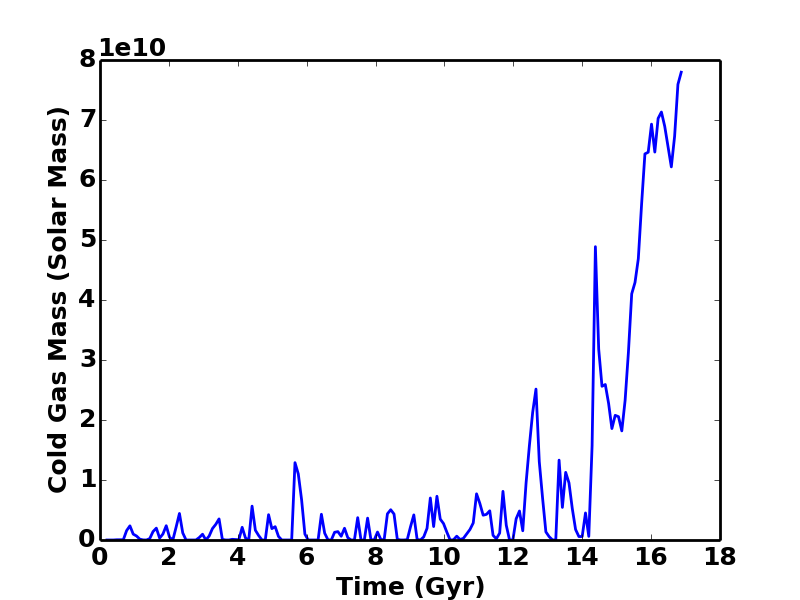}
\caption{The total amount of cold gas in the central 100 kpc of the cluster as a function of time in the simulations with low resolution. Top: $l_{max}=9$ and $\Delta x_{min} \sim 500$ pc. Bottom: $l_{max}=8$ and $\Delta x_{min} \sim 1$ kpc.
\label{fig:low_resolution}}
\end{center}
\end{figure}

The lowest resolution run with $l_{max} = 8$ experiences even more cycles of AGN outbursts before a massive stable disk finally forms at $t\sim 14$ Gyr. The time evolution of the total amount of cold gas in the central 100 kpc of the cluster in the two low resolution simulations is shown in Figure ~\ref{fig:low_resolution}.

Overall, we find that simulations with higher resolution produce more realistic looking cold clumps and filaments, but as long as the resolution is above $l_{max} = 10$, the long-term evolution of the cluster stays the same: the cluster experiences four phases ending with a large cold disk described in detail in Section~\ref{sec:results_evolution}. When the resolution is lower than $l_{max} = 10$, the cold clumps and the cold disk are only resolved by a few cells. Somewhat to our surprise, this results in cycles of AGN outbursts and clump formation that seem to agree with the observations better than our standard run. We will discuss more on this in Section~\ref{sec:discussion}. 

\subsubsection{Kinetic Fraction}\label{sec:results_f}

As described in Section~\ref{sec:methodology_jet}, we assume that half of the jet energy is in the form of kinetic energy with the rest in thermal (i.e. $f_{\rm kinetic} = 0.5$) in our standard simulation and the previous resolution test runs. However, the choice of the parameter $f_{\rm kinetic} = 0.5$ is not physically motivated, because we do not resolve the actually base of the AGN jets in our simulations, which is only a few Schwarzschild radii from the SMBH \citep{JetBase}, and it is unclear exactly how much energy is thermalized at the location of our jet launching planes. Therefore, we carry out two simulations in two extreme cases to test the influence of the choice of this parameter: one with pure kinetic feedback ($f_{\rm kinetic} = 1.0$) and one with mostly thermal feedback ($f_{\rm kinetic} = 0.1$). 

With pure kinetic feedback, the jet material has even lower pressure and the jets are narrower. The clumps initially form at slightly larger radii owing to the faster velocity of the jets as we have discussed in Paper II. The rest of the cluster evolution is very similar to our standard run.  

The result from the simulation with mostly thermal feedback also agrees with the standard run with only a subtle difference. Due to the lower velocity of the jets, most clumps form at radii slightly smaller than in the standard run with a lower initial velocity. Thus the average specific angular momentum of the cold gas is smaller. The total amount of the gas that cools during the clump formation stage is comparable to the standard run, but the total angular momentum of the cold gas measured at $t\approx1$ Gyr is only half of that in the standard simulation. This results in a slightly larger amount of cold gas accreted onto the SMBH at early times, and therefore a stronger heating from the AGN. The cooling rate of the ICM is further reduced by $50\%$ compared with our standard simulation. 

Overall, we find the exact choice of the kinetic fraction $f_{\rm kinetic}$ to have little effect on the long term thermal balance of the cluster, i.e. AGN is able to balance cooling with a wide range of $f_{\rm kinetic}$ (from $10\%$ to $100\%$). 

\subsubsection{Feedback Efficiency}\label{sec:results_epsilon}

In all the simulations discussed previously, we have been using a rather moderate feedback efficiency $\epsilon$ of $0.1 \%$. However, the estimation of $\epsilon$ can be as high as $10 \%$ in some systems\citep[e.g.][]{Churazov2005}. To study the effect of a higher $\epsilon$, we carry out a simulation with $\epsilon=1\%$ at medium resolution. 

\begin{figure}
\begin{center}
\includegraphics[scale=.39,trim=0cm 0cm 0.4cm 0cm, clip=true]{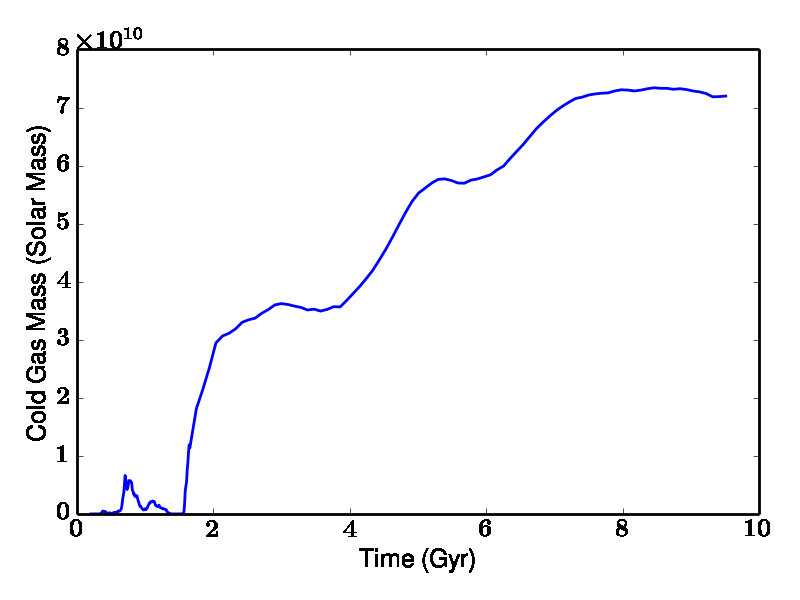}\\
\includegraphics[scale=.4,trim=0.4cm 0cm -0.2cm 0cm, clip=true]{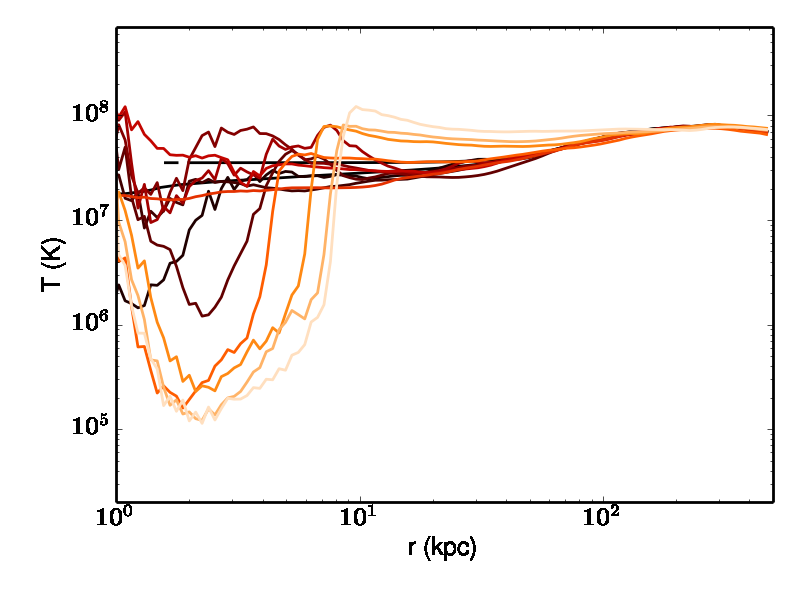}
\caption{Top: the total amount of cold gas in the central 100 kpc of the cluster as a function of time in the simulation with high feedback efficiency ($\epsilon = 1\%$). Bottom: the evolution of the temperature profile of the gas weighted by cell mass. At late times (light yellow lines), the temperature profile becomes flat and the cluster is turned into a non cool-core cluster. 
\label{fig:high_epsilon}}
\end{center}
\end{figure}

Due to the higher $\epsilon$, the AGN power is higher when cooling first starts to run away. Thus the initial burst of clump formation at $t\sim 1$ Gyr is quickly suppressed, with the peak amount of cold gas an order of magnitude lower than our standard run. AGN is turned off when all the cold gas is accreted, which allows the ICM to cool again. The cluster experiences a larger burst of clump formation at $t\sim 2$ Gyr which results in the formation of a large stable disk. The formation of the stable disk is delayed by 1 Gyr compared with our standard simulation. The rest of the cluster evolution is similar to that in our standard run, except that at $t\sim$ 4 and 6 Gyr, the cluster experiences two short episodes of clump formation at smaller radii (of $\sim 10$ kpc) than the initial burst, which correspond to the two jumps in the amount of cold gas in the top panel of Figure~\ref{fig:high_epsilon}. 

Overall, cooling is suppressed by the AGN feedback successfully in this simulation with high efficiency. However, the temperature profile of the gas becomes flat inside the cluster core at late times (see bottom panel of Figure~\ref{fig:high_epsilon}) due to the slightly excessive energy output from the AGN, practically turning the cluster into a non-cool-core cluster. Since AGNs are mostly found in the center of cool-core clusters, we consider this model acceptable but less favorable than our standard simulation with $\epsilon=0.1\%$.

\section{Discussion}\label{sec:discussion}
Here we discuss how our results compare with previous simulation works in Section~\ref{sec:discussion_sim} and with the observations of nearby cool-core clusters in Section~\ref{sec:discussion_obs}. We have analyzed how AGN feedback successfully suppresses cooling in our simulations, but our model is not perfect. We discuss the major issues and possible solutions in Section~\ref{sec:discussion_problems}.

\subsection{Comparison with Previous Simulations}\label{sec:discussion_sim}

We find good agreement between our results and \citet{Gaspari2012} in general. This is not surprising given that our jet modelings are similar and the resolution in \citet{Gaspari2012} is close to our medium resolution simulation with $l_{max}=10$ which shows converging results as our standard run. The feedback efficiency in our standard simulation ($\epsilon=0.1\%$) is roughly an order of magnitude lower than theirs ($\epsilon = 0.6 \%$ and $1\%$), but the general agreement between our results, as well as our test run with $\epsilon=1\%$ (discussed in Section~\ref{sec:results_epsilon}) shows that the results are not very sensitive to the choice of the feedback efficiency $\epsilon$. Both our work here and \citet{Gaspari2012} find that in a momentum-driven AGN feedback model, a thermal balance can be achieved with a wide range of feedback efficiency.  

We also find that in both works, spatially extended cold gas is seen mostly at early times, while at late times, over a few Gyr, cold gas exists in the form of the rotating structure in the center of the cluster. The final amount of the cold gas is of order $10^{11}$ M$_{\odot}$ in both works.

\begin{figure}
\begin{center}
\includegraphics[scale=.4,trim=0cm 0cm 0.3cm 0cm, clip=true]{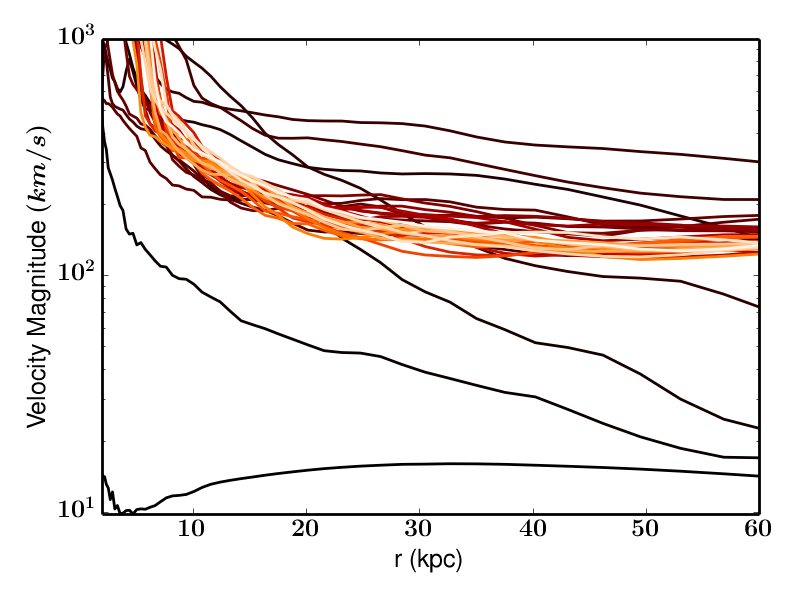}
\includegraphics[scale=.41,trim=0.2cm 0cm 0cm 0cm, clip=true]{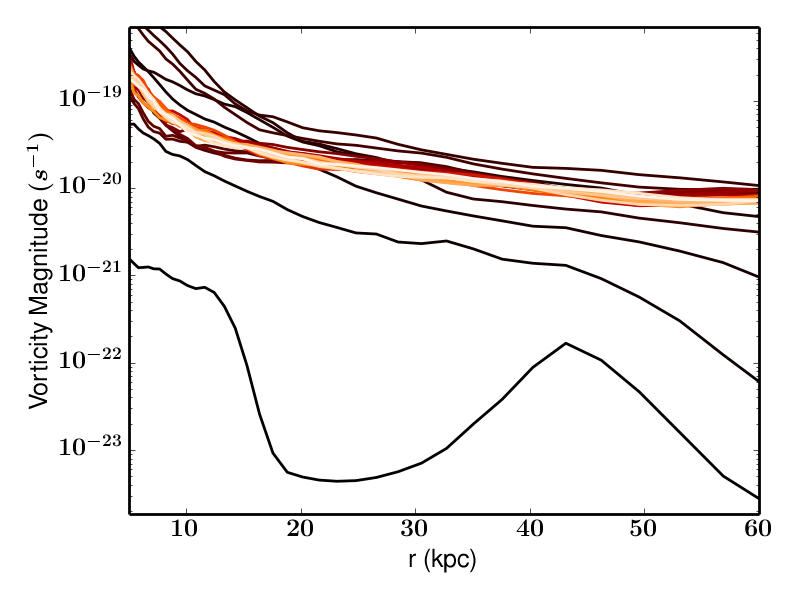}
\caption{The profile of volume-weighted velocity magnitude (top) and vorticity magnitude (bottom) at different times, sampled every 20 Myr. These are rough measurements of the level of turbulence. At very early times, they are both low (dark brown). They increase as the jet power increases and settle to a steady profile in the last stage of evolution. 
\label{fig:sigma}}
\end{center}
\end{figure}

The critical ratio of the thermal instability timescale over the free-fall time ($t_{TI}/t_{ff}$, or the cooling time over the dynamical time $t_{cool}/t_{dyn}$) for the ICM to condense into a multiphase medium in our simulations is generally consistent with that found in \citet{Sharma12} and \citet{Gaspari2012} except at very early times. As we have discussed in Paper II, we find a critical ratio of $\sim 3$ for the ICM to first cool into clumps. This ratio is small likely because turbulence in our simulations is solely driven by the AGN jets, and is therefore weak at early times when the feedback power is still low. Figure~\ref{fig:sigma} shows how the velocity and vorticity magnitude (as indicators of the turbulent level) of the gas evolve with time. The level of turbulence is low initially, but increases quickly with time as the jet power increases. After $t\sim 0.5$ Gyr, global turbulence has set up and clump formation is seen only when the $t_{TI}/t_{ff}$ ratio is below $\sim 10$, consistent with the critical value found in \citet{Sharma12} and \citet{Gaspari2012}. In our standard run, clump formation stops once the cluster enters the last phase with the stable disk, during which time the ratio is always over 10. 

Compared with \citet{Cattaneo2007}, we find that even though the feedback modeling differs and our feedback efficiency $\epsilon$ is only $1\%$ of theirs, in both simulations, the cluster experiences a large AGN outburst at early times and the system later settles to a long-lived quiescent phase. The total amount of cold gas in the last stage is larger in our simulations likely because our model cluster is much more massive than that in \citet{Cattaneo2007}.

\citet{Gaspari2011} studied the momentum driven AGN feedback with mass-loaded outflows, and also found that the results are not very sensitive to the efficiency $\epsilon$ in their cold-regulated feedback model.

\citet{Reynolds06} finds that simple hydro jets cannot prevent a cooling catastrophe because most of the energy channels out of the cool core. With small-angle jet precession in our model, we find that even though the cooling rate shoots up to $10^3$ M$_{\odot}/\rm yr$ at $t\sim 1$ Gyr (Figure~\ref{fig:Mdot}), it quickly declines following the AGN outburst. More than half of the feedback energy still channels out of the core (Figure~\ref{fig:heating} and Section~\ref{sec:results_heating_cooling}), but a classic cooling flow is prevented.

In the cosmological simulation in \citet{Dubois2010}, momentum driven AGN feedback is also found to successfully suppress cooling in the center of the cluster, which results in a smaller cooling flow and a reduced stellar mass of the BCG.

\subsection{Comparison with Observations}\label{sec:discussion_obs}

\begin{figure*}
\begin{center}
\includegraphics[scale=.4]{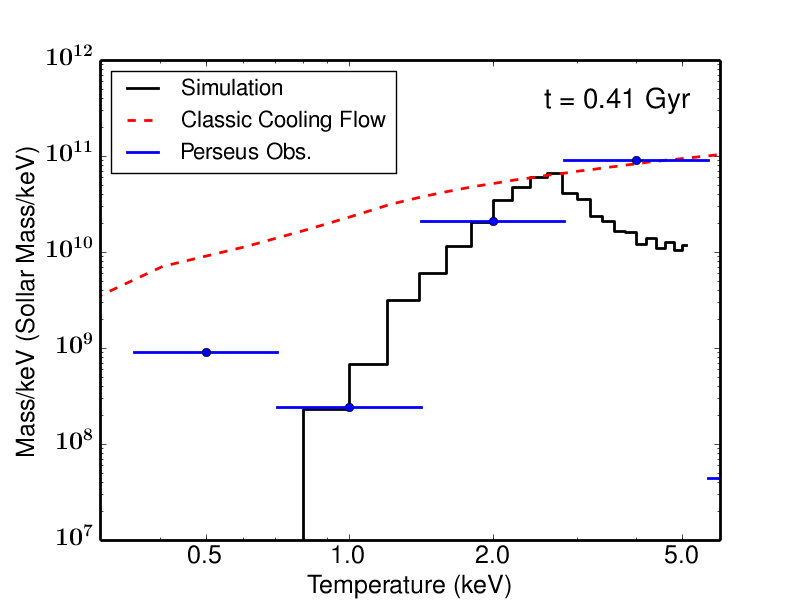}
\includegraphics[scale=.4]{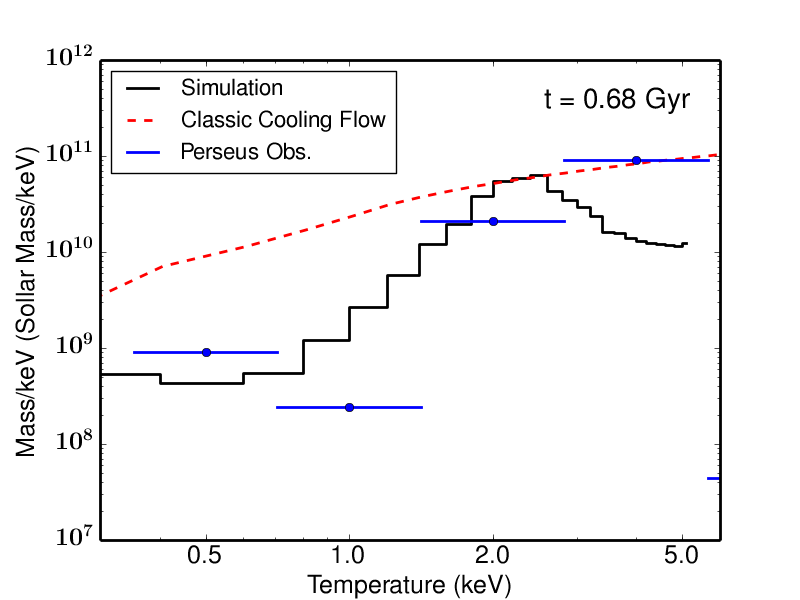}\\
\includegraphics[scale=.4]{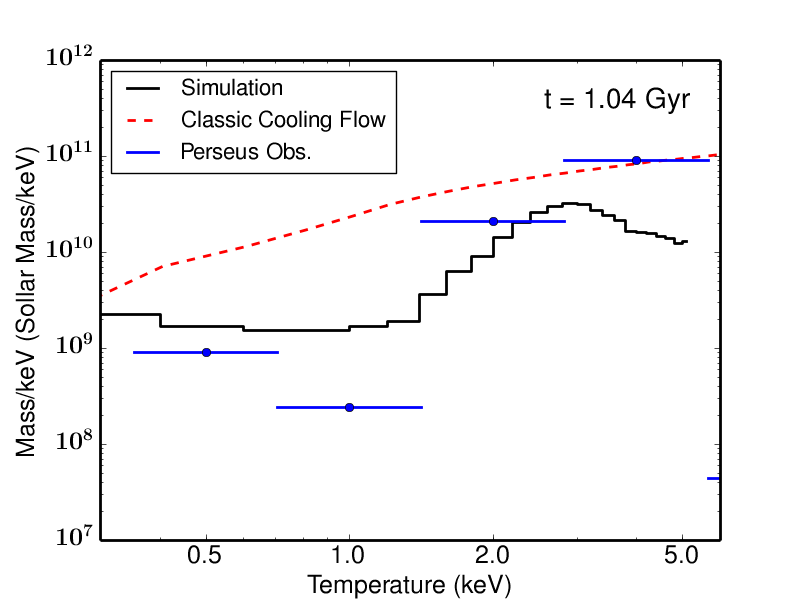}
\includegraphics[scale=.4]{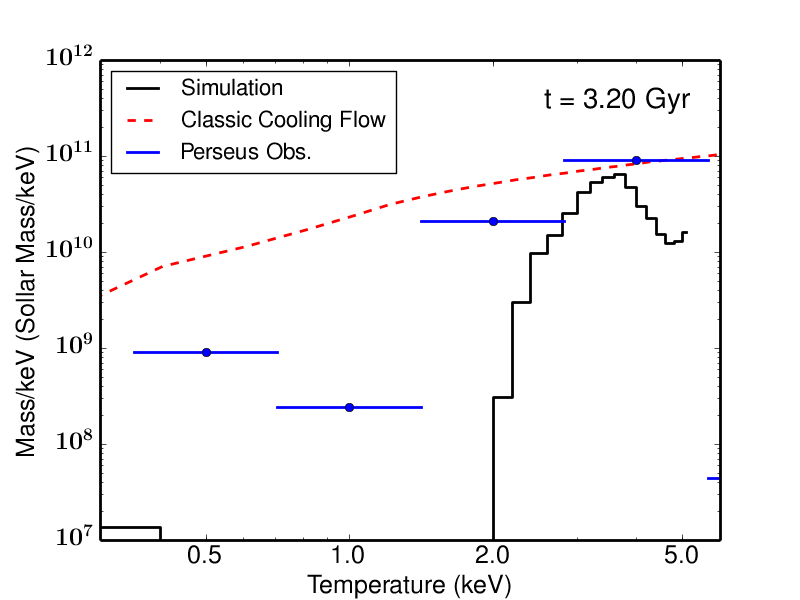}
\caption{Distribution of the total mass of the gas at fixed temperature bins within $r< 32$ kpc of the cluster center at different stages of our standard simulation. The black solid line shows the result from our simulation with a bin size of 0.2 keV. The blue dots are the observations of Perseus within the innermost 1.5 arcmin ($\sim 32$ kpc) and the red dashed line is the classic cooling flow model prediction taken from \citet{Fabian2006}. The lack of gas at $\sim 1-2$ keV in our simulation is consistent with the observations. 
\label{fig:M_T}}
\end{center}
\end{figure*}

We have shown in Section~\ref{sec:results_heating_cooling} that a classic cooling flow is prevented and the cooling rate in our standard simulation is reduced to $\sim 30$ M$_\odot/$yr, consistent with the estimation for the Perseus cluster \citep{Fabian2006}. More directly, we can compare the mass-temperature distribution of the gas in our simulation with Figure 11 of \citet{Fabian2006}, where they fit the $Chandra$ X-ray spectra with a multi-temperature model and find the total mass of the gas within each temperature bin. This comparison is shown in Figure~\ref{fig:M_T} at four different times from the four stages corresponding to those in Figure~\ref{fig:projections}. In all stages, the amount of gas around $1-2$ keV in our simulation is significantly lower than the classic cooling flow prediction and generally agrees with the observations of the Perseus cluster. The recovery at around $0.5$ keV is missing in the first panel (the initial cooling phase) and the last panel (the disk phase), but is seen in the middle two panels when the ICM is cooling into extended structures that morphologically resemble the $\rm H\alpha$ filaments observed in Perseus. This confirms that the $0.5$ keV gas and the filaments are closely related.

\begin{figure*}
\begin{center}
\includegraphics[scale=.575,trim=2.5cm 1cm 2.5cm 1.0cm, clip=true]{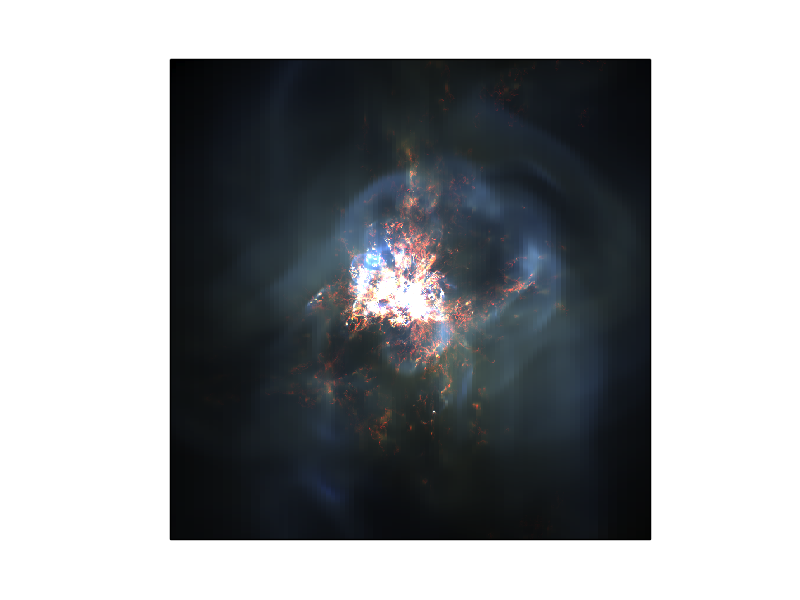}
\includegraphics[scale=.51,trim=1.5cm 0.2cm 1.0cm 0cm, clip=true]{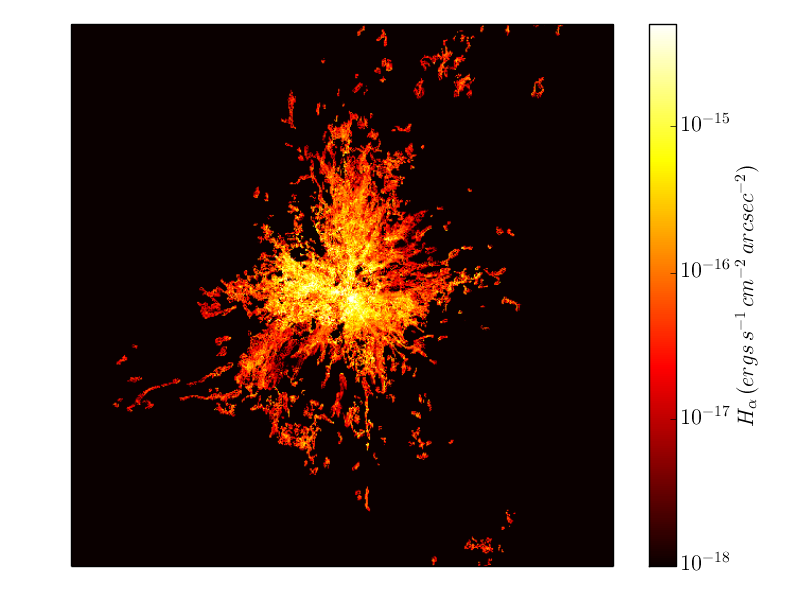}
\caption{Left: the synthetic X-ray composite image of the central $r<50$ kpc region of the simulated cluster at $t3=1.04$ Gyr in 0.3-1.2 (red), 1.2-2 (green) and 2-7 keV (blue) bands. The observational angle is randomly chosen to be $23^{\circ}$ from the z-axis in the y-z plane. The size of the region is comparable to the Chandra X-ray map of Perseus in \citet{Fabian2006}. Each individual channel has been rescaled to bring out fainter features. The ``bubbles'' seen in the simulations are created by shock waves and the enhanced soft X-ray (red) traces the cold clumps and filaments. Right: the synthetic $\rm H\alpha$ map of the same region from the same angle. The $\rm H\alpha$ filaments spatially correlate with the enhanced soft X-ray, but the former shows more detailed structures.
\label{fig:X}}
\end{center}
\end{figure*}

The correlation between the soft X-ray gas ($0.5-3$ keV) and the cold filaments is also be seen in the composite X-ray image of Perseus (e.g. Figure 8 of \citet{Fabian2003X} or Figure 3 of \citet{Fabian2006}). We have already shown in Paper II that the $\rm H\alpha$ and the soft X-ray maps correlate with each other, consistent with the observations. This is because both $\rm H\alpha$ and soft X-ray are enhanced in the thin transition layer surrounding the cold clumps and filaments, and in the stripped tails behind the cold gas. We present the synthetic composite X-ray image and the $\rm H\alpha$ map from our standard simulation at $t_3=1.04$ Gyr in Figure~\ref{fig:X}. Like in the X-ray image of Perseus, the hard X-ray captures the shocks whereas the filaments stand out in the soft X-ray (red) that spatially correlates with the $\rm H\alpha$ emitting gas.

\begin{figure}
\begin{center}
\includegraphics[scale=.45]{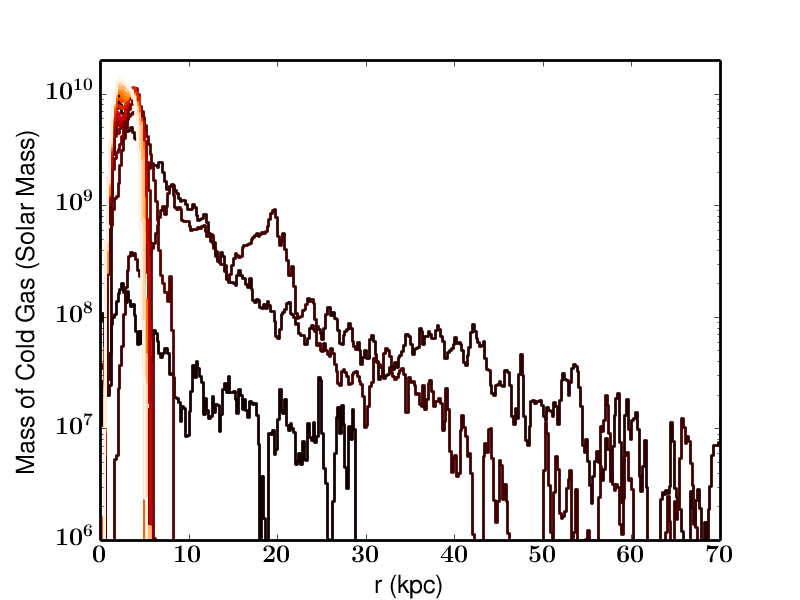}
\caption{The radial distribution of cold gas at different times in our standard simulation, sampled every 20 Myr as in Figure~\ref{fig:profile}. The bin size is $0.2$ kpc. Extended cold gas is only seen at early times (dark lines), and at later times all the cold gas is inside the disk of about 6 kpc.
\label{fig:M_r}}
\end{center}
\end{figure}

When comparing the spatial distribution of the cold gas in our standard simulation with the observations, we find that the cold gas is always confined within the cooling radius $R_{cool}$, the radius at which the gas cooling time is 5 Gyr, consistent with the observational constraint found in \citet{McDonald10} as we have discussed in Paper II and is shown in Figure~\ref{fig:M_r}. 

The structure of the cold gas (formed along the propagation direction of the jet) in the clump formation stage in our simulation morphologically resembles the elongated molecular gas seen in some cool-core clusters such as Abell 1795 \citep[e.g.][]{McDonald2009}, while the extended filamentary structure (see Figure~\ref{fig:projections} and Figure~\ref{fig:X}) in the third stage looks similar to that found in Perseus and some other clusters (e.g. Abell 0496 in \citet{McDonald10}). The total mass of the cold gas in our simulation ($\sim 10^{11}$ M$_{\odot}$) is higher than but still consistent with the lower limit for Perseus ($\sim 4\times 10^{10}$ M$_{\odot}$ in \citet{Salome2006}). However, this much cold gas in the form of a rotating disk in the final stage of our simulations lacks observational support. Even though many cool core clusters are seen to harbor a compact rotating structure in the center, the observed disks are usually orders of magnitude less massive than that in our simulation. For example, \citet{Hamer2014} found a disk of $\sim 5$ kpc in Hydra-A, similar to our disk in radius, but its mass is estimated to be only $2.3 \pm 0.3 \times 10^9$ M$_{\odot}$. Other observed disks are either of similar mass or even smaller \citep[e.g.][]{Lim2000, ColdDisk2010}.  

The disk in our simulation also stays stable for a few Gyr, which would predict most cool-core clusters to be in this stable phase. However, only a fraction of the cool-core clusters in the survey of \citet{McDonald10} host a compact central structure. Therefore, we consider the last phase of our simulation unrealistic and thus additional physics is still needed.

\subsection{Problems of our Model and Ideas for Solution}\label{sec:discussion_problems}

As discussed in Section~\ref{sec:discussion_obs}, the major problem with our model is that the standard simulation produces a cold disk that is too large and stays too long compared with observations. Interestingly, this problem is most serious in the high and medium resolution simulations. In the low resolution simulations, the cluster experiences cycles of clump/filament formation with short-lived disks, which is more consistent with the expectation from the observations. Even though a large disk eventually forms, in the lowest resolution run with $l_{max}=8$, the disk forms after 14 Gyr, which is roughly the age of the universe. Therefore, the low resolution probably has the same effect as the crucial physics that is not included in our model. 

Two major effects of low resolution are: first, cold clumps are harder to form and more easily destroyed; second, a cold disk is more quickly accreted onto the SMBH and is thus destroyed. The physics missing in our simulations includes thermal conduction, star formation, self-gravity, magnetic fields, and viscosity.

Heat conduction alone is unable to prevent a classic cooling flow \citep[e.g.][]{Narayan2003, P1}, but it does affect the temperature of the gas with cooling time of a few Gyr at $\sim 100$ kpc, and is found to produce a more homogeneous ICM with less temperature variation in the cluster core in cosmological simulations \citep{Dolag2004, Britton2013}. By transporting thermal energy from larger radii inward, heat conduction may reduce the required amount of heating from the AGN. \citet{Wagh2013} found that anisotropic conduction does not change the criteria for gas to cool into multiphase medium, but it might affect the amount of the gas that cools into clumps and filaments, and reduce the disk mass at later times.

Star formation has been observed in many cool core clusters, with a typical rate of $1-10\%$ of the mass deposition rates estimated from the X-ray assuming a steady classic cooling flow \citep{Hicks2005, ODea2008, ODea2010, Hicks2010}. Even star bursts of much higher rates are occasionally detected \citep[e.g.][]{Phenix2012}. Star formation does not only consume the cold gas, the feedback from stars can also locally heat up the cold gas. Given the typical observed star formation rate of $\sim 10$ M$_{\odot}$/yr in the center of cool-core clusters, the rotating disk in our simulation should be completely consumed by star formation alone within a few Gyr, which may allow the ICM to cool and experience multiple cycles of cold filament formation as seen in the low resolution simulations. The feedback from star formation may also help drive the formation of filaments \citep{Falceta2010}. We would like to explore the effects of star formation in the near future.

Self-gravity may trigger instabilities inside the cold disk that can drive inflows, which may trigger accretion onto the SMBH, reducing the disk mass.

Magnetic fields will likely make the cold clumps more filamentary as we discussed in Paper II and is shown in \citet{Sharma2010} and \citet{Wagh2013}. They can also affect how heat is transported via conduction, especially around the cold filaments, and support them against gravity .

As we have mentioned earlier, we do not have explicit viscosity in our simulations. Including viscosity can increase the accretion rate of the cold disk and thus reduce its mass. Viscosity also helps dissipate turbulent energy \citep{Ruszkowski2004, Ruszkowski2005, Zhu2013}, which increases the efficiency of heating. 

All the physics discussed above has the potential to alleviate the problem in our simulation, and may also have an impact on the filamentary cold gas, which we leave for future studies.

\section{Conclusion}\label{sec:conclusion}
In this study, we have performed a series of three-dimensional AMR simulations examining the effect of radio-mode AGN feedback in an idealized cool-core galaxy cluster built to match the observations of the Perseus cluster. The key question we are trying to address here is whether AGN feedback can prevent a classic cooling flow in a realistic fashion over a long period of time. We model the momentum driven AGN feedback with a pair of jets precessing at a small angle, and we compute the outflow rate and the jet power based on the accretion rate of the cold gas surrounding the SMBH. Our standard simulation has a minimum cell-size of $\sim 60$ pc, and the cluster evolves for 7 Gyr. Our primary results are summarized as follows.

(1) Run-away cooling first occurs only in the very center of the cluster at $t\sim 200$ Myr while no local instability develops outside the cooling region. The cold gas is accreted onto the super-massive black hole which powers AGN jets at an increasing rate as the entropy of the ICM continues to decrease in the cluster core. At $t\sim 450$ Myr, cold clumps of gas first start to form at $10-20$ kpc along the propagation direction of the AGN jet due to its non-linear perturbation. The cold clumps fall to the center of the cluster roughly within a dynamical time. Some of the cold gas blocks and redirects the jets to go out in other directions, causing clumps to form in all directions within the cluster core, while some feeds the SMBH, increasing its power which causes the core entropy to increase and thus slows down the cooling of the ICM. Due to its non-zero (random) angular momentum, not all the gas can be accreted onto the SMBH. By $t\sim 1.7$ Gyr, all the cold gas has settled to a rotationally supported disk of $\sim 6$ kpc in radius and $\sim 10^{11}$ M$_{\odot}$ in mass. The cold disk stays, keeping the AGN on for a few Gyr through the end of the simulation.

(2) The AGN heating rate traces the ICM cooling rate of the cluster core. About $30\%$ of the energy goes to balance cooling within $r<100$ kpc, with the rest deposited outside of the cluster core. The mass cooling rate only briefly reaches the classic cooling flow rate at the peak of the cold clump accretion phase; the average cooling rate is only $30$ M$_{\odot}/$yr, an order of magnitude lower than the classic cooling flow value. Overall, the ICM cooling is well balanced by AGN heating, and a cool-core appearance of the cluster is also preserved.

(3) The long term evolution of the cluster is not sensitive to the resolution of the simulation either, as long as it is better than $\Delta x_{min}\approx 0.5$ kpc. When $\Delta x_{min} \approx 0.5$ kpc, because of the smaller amount of the cold clumps and quicker accretion of the cold disk, the cold gas disk can be completely accreted and the cluster experiences cycles of clump formation followed by AGN outbursts. The cold massive disk still forms eventually after a few Gyr. With even coarser resolution of $\Delta x_{min} \approx 1$ kpc,, the period of the cycle gets shorter and the formation of the final stable disk is delayed even more (longer than the age of the universe). 

(4) Our parameter study shows that the general results are not very sensitive to the model parameters including the fraction of kinetic energy in the jet $f_{\rm kinetic}$, and the feedback efficiency $\epsilon$, with only subtle differences in the details. The self-regulating model enables AGN feedback to balance cooling with a wide range of parameters ($f_{\rm kinetic}$ from 0.1 to 1, and $\epsilon$ from $0.1\%$ to $1\%$).

(5) Our results are generally consistent with previous simulations \citep[e.g.][]{Gaspari2012}. Besides preventing a classic cooling flow, our standard simulation also successfully produces spatially extended structures that bear striking similarity in both the morphology and the spatial extension to the line-emitting filamentary gas observed in nearby cool-core clusters. However, the cold disk in the final stage of our simulation is much more massive than the compact rotating structures that have ever been observed in cool-core clusters, and it stays for too long, which contradicts the observed frequency of such structures. This implies that our model, while successful in many aspects, is still missing important physical effects, such as thermal conduction, star formation, self-gravity, magnetic fields, and viscosity. We plan to incorporate them to study their effects in the future.

\acknowledgments

We thank A. J. R. Sanderson for providing the observational data in Figure~\ref{fig:M_T}. We also thank Arif Babul, Mark Voit, Brian O'Shea and Greg Meece for useful discussions. We acknowledge financial support from NSF grants AST-0908390, AST-1008134, AST-1210890 and NASA grant NNX12AH41G,  as well as computational resources from NSF XSEDE and Columbia University.


\end{document}